\documentclass{aa}  

\usepackage{graphicx}
\usepackage{txfonts}
\usepackage{bm}
\usepackage{amsmath}
\usepackage{xcolor}
\usepackage{hyperref}
\hypersetup{
    colorlinks,
    linkcolor={red!50!black},
    citecolor={blue!50!black},
    urlcolor={blue!80!black}
}
\usepackage{cleveref}
\usepackage{tikz}
\usetikzlibrary {positioning, shapes.geometric} 
\usepackage{pdflscape}

\usepackage{layouts}

\let\originalleft\left
\let\originalright\right
\renewcommand{\left}{\mathopen{}\mathclose\bgroup\originalleft}
\renewcommand{\right}{\aftergroup\egroup\originalright}

\newcommand{\eg}{\emph{e.g.}}
\newcommand{\ie}{\emph{i.e.}}
\newcommand{\resp}{\emph{resp.}}

\newcommand{\Hamlet}{{\scshape Hamlet}}
\newcommand{\pmwd}{{\scshape pmwd}}
\newcommand{\HamletPM}{{\scshape Hamlet-PM}}
\newcommand{\LCDM}{$\Lambda$CDM}

\newcommand{\kms}{\,{\rm km/s}}
\newcommand{\mg}{\,{\rm mag}}
\newcommand{\Mpc}{\,{\rm Mpc}}
\newcommand{\Mpch}{\,h^{-1}\Mpc}
\newcommand{\hMpc}{\,h\Mpc^{-1}}
\newcommand{\Msh}{\,h^{-1} {\rm M}_\odot}

\newcommand{\lD}{\mathcal{D}}
\newcommand{\lH}{\mathcal{H}}
\newcommand{\lM}{\mathcal{M}}
\newcommand{\lN}{\mathcal{N}}
\newcommand{\lO}{\mathcal{O}}
\newcommand{\lP}{\mathcal{P}}
\newcommand{\lZ}{\mathcal{Z}}

\newcommand{\muo}{\mu^{\rm obs}}
\newcommand{\zo}{z^{\rm obs}}

\newcommand{\vk}{\bm{k}}
\newcommand{\vr}{\bm{r}}
\newcommand{\vv}{\bm{v}}
\newcommand{\vnab}{\bm{\nabla}}
\newcommand{\hr}{\bm{\hat{r}}}

\newcommand{\rhocic}{\rho^{\rm CiC}}
\newcommand{\vcic}{\vv^{\rm CiC}}
\newcommand{\deltacic}{\delta^{\rm CiC}}
\newcommand{\vlin}{\vv^{\rm lin}}
\newcommand{\pval}{$p$-value}
\newcommand{\sig}[1]{$#1\sigma$}

\defcitealias{Pfeifer2023}{P23}
\defcitealias{McAlpine2025}{MA25}
\defcitealias{Hernandez-Martinez2024}{HM24}
\newcommand{\cpa}[1]{\citepalias{#1}}
\newcommand{\cta}[1]{\citetalias{#1}}

\begin{document} 

   \title{Constraining cosmological simulations with peculiar velocities: a forward-modeling approach}

   \titlerunning{Constraining simulations with velocities}

   \author{A. Valade \inst{1,2,3}, N. Libeskind \inst{2}, D. Pomarède \inst{4}, R. Stiskalek \inst{5}, Y. Hoffman \inst{3}, S. Gottlöber \inst{2} \and R. B. Tully \inst{6}} 

   \institute{
            Aix Marseille Universit\'e, CNRS/IN2P3, CPPM, Marseille, France \\
            \email{avalade@aip.de} 
        \and
            Leibniz-Institut für Astrophysik Potsdam,
            An der Sternwarte 16, 14482 Potsdam, Germany 
        \and 
            Racah Institute of Physics,  
            Hebrew University, Jerusalem 91904, Israel 
        \and
            Institut de Recherche sur les Lois Fondamentales de l’Univers, 
            CEA, Universite’ Paris-Saclay, 91191 Gif-sur-Yvette, France
        \and
            Astrophysics, University of Oxford, 
            Denys Wilkinson Building, Keble Road, Oxford, OX1 3RH, UK 
        \and
            Institute for Astronomy, University of Hawaii, 2680 Woodlawn Drive, Honolulu, HI 96822, USA}

   \date{Received Month XX, 20XX; accepted Month XX, 20XX}

 
  \abstract
   {Numerical simulations are a key tool to decipher the dynamics of gravitation. Yet, they fail to spatially reproduce the Universe we observe, limiting comparison between observations and simulations to a statistical level. This is highly problematic for rare, faint or well studied nearby objects that are observed in a single environment. The computational cost of recovering this environment in random simulations is prohibitive for most realistic applications.}
   {We present \HamletPM, a method that enables the constraining of initial conditions for cosmological simulations so as to produce evolved numerical universes that can be directly compared to observations of the Local Universe. Such simulations are known as \emph{constrained} simulations.}
   {Our method implements the field-level forward modeling of the early-time density field from sparse and noisy measurements of late-time peculiar velocities. The dynamics are integrated with a coarse particle-mesh gravity solver, thus probing the mildly non-linear regime. We apply our code to the Cosmicflows-4 compilation of peculiar velocities up to $z\sim0.05$ ($160\,h^{-1}{\rm Mpc}$). The resulting constrained simulations are analyzed under the framework of the Local Universe Model as well as an amended version that captures systematic spatial shifts in the scattering of simulated counterparts.}
 {A series of one hundred dark-matter only cosmological simulations with a resolution of $512^3$ particles in a $500^3\,[h^{-1}{\rm Mpc}]^3$ box are presented. These spatially reproduce the observed large-scale structure in the constrained volume. Special attention is given to twelve prominent nearby galaxy clusters, whose simulated counterparts are matched on criteria of mass and separation. We provide a mass estimate constrained by the dynamical environment for each cluster.} 
   {Field-level forward modeling of the initial conditions can be applied to actual observations so as to produce highly constrained cosmological simulations. In its current form, this method already overtakes in quality the pipeline currently in use in the peculiar-velocity community, although systematic biases still need to be addressed. Furthermore, improving the model can be done easily thanks to the inherent flexibility of the Bayesian approach.}

   \maketitle


\section{Introduction}

 
The large-scale distribution of matter in the Universe is  understood to be the result of initial quantum fluctuations imprinted during the inflationary stage of the big bang followed by billions of years of gravitational instability. Although complex, it is a numerically tractable problem given the initial over- and under-density field, and the relatively simple law of gravitational attraction. The nature of the primordial perturbations in the density and velocity field is well established via observations of the temperature fluctuations in the cosmic microwave background. The CMB \citep[\eg{}][]{PlanckCollaboration2020} describes these perturbations statistically, allowing the large scale structure to be simulated in a statistical sense, namely with the goal of producing representative volumes of the Universe that are consistent with having grown out of the primordial power spectrum of fluctuations.

Such simulations are indispensable tools to understand the framework within which galaxies form and have been used extensively. Indeed a large variety of codes have been developed to model the complex hydrodynamics, star formation and feedback processes with enormous success \citep[for a recent review, see ][]{Vogelsberger2019}. Yet, while important in modeling the galaxy population as a whole, such simulations cannot be as easily  compared with the single individual realization of the universe that we inhabit, namely the ``Local Universe''. To constrain models from observations, cosmologists then resort to summary statistics, such as the two-point correlation \citep[\eg{}][]{Adame2025} function or the Alcock-Paczyński test \citep[\eg{}][]{Cuceu2025}. 

This paper is part of the effort to use low redshift observations of the Local Universe to produce spatially \emph{constrained} initial conditions, which, once evolved, result in \emph{constrained} simulations that can be directly compared to the observations \citep{Kravtsov2002, Klypin2003}. 

One of the main uses of such simulations so far is the study of galaxy clusters, such as Virgo, Coma, Perseus, etc. Quantities of interest include their mass \citep[\eg{}][hereafter P23]{Pfeifer2023}, merging history and formation \citep[\eg{}][]{Sorce2016, Sorce2020}, connectivity \citep[hereafter MA25]{McAlpine2025}, dynamical signature along the line of sight \citep{Sorce2024}, abundance \citep{Dolag2023} and various other observables \citep[X-Ray luminosity, temperature and SZ effect; ][hereafter HM24]{Hernandez-Martinez2024}. Constrained simulations have also been utilized to constrain the galaxy bias \citep{Hoffman2018}.

Two main approaches to construct such initial conditions have been investigated over the last decade, each relying on different data sources, namely redshift data and peculiar velocity data. Redshift surveys have paved the way for the discovery and exploration of the large scale structure, from the 1980s \citep{DeLapparent1986} until today \citep[\eg{}][]{DESICollaboration2025}. Modern surveys are dense, precise (with redshift errors of $\sigma_z \leq 10^{-4}$) and tend towards completeness in increasing volumes. Yet, their use for the problem of reconstructing the underlying density field is made difficult by galaxy bias \citep[for a review]{Desjacques2018}. The modeling of this relationship on non-linear scales is complex and thus has to be approximated by empirical laws to maintain computational tractability in the context of field reconstruction \citep{Jasche2019, McAlpine2025}. 

The redshift contains information about both the expansion of the Universe and the local movement of a galaxy in its environment, known as the peculiar velocity. If measurements of physical distances are available, individual peculiar velocities can be extracted from the redshift data. This allows for the circumvention of galaxy bias since peculiar velocities are sensitive to the total distribution of matter, dark as well as luminous. Yet, measurements of distances are difficult to obtain and catalogs thereof are thus scarcer, noisier and prone to interpretation biases.

The Constrained Local UniversE Simulation consortium \citep[CLUES;][]{Yepes2009, Doumler2013} has pioneered the use of peculiar velocities for field reconstructions and constraining simulations \citep[\eg{}][]{Sorce2014, Sorce2020, Sorce2024, Hoffman2018, Libeskind2020, Pfeifer2023, Dolag2023}. This pipeline unfolds in three steps: (1) the reconstruction of the density and velocity fields from observational data with a Wiener-Filter \citep[WF;][]{Zaroubi1995}; (2) the shifting of the constraints back in time with a Reverse Zeldovich Approximation \citep[RZA;][]{Doumler2013A}; (3) the creation of initial conditions for Constrained Simulations \citep[CRs;][]{Hoffman1992, Doumler2013c}. 
This approach is somewhat limited by the inherent approximations of each step. The Wiener Filter assumes a linear relationship between the density and peculiar velocity field. While this is an acceptable assumption on large scales (beyond a few Mpc), it breaks down in the non-linear regime. The method also assumes that the uncertainties on the velocity are normally distributed. The latter hypothesis is at odds with the reality of velocity data, which motivated the development of ``un-biasing'' techniques \citep[\eg{}][]{Sorce2015, Hoffman2021, Sorce2023}. 

The application of field-level forward modeling to peculiar velocities to reconstruct the velocity field in the late-time Universe, was first proposed by \citet{Lavaux2016} and then later applied to the Cosmicflows-3 catalog \citep{Tully2016} by \citet{Graziani2019}. \citet{Boruah2020} and \citet[\Hamlet]{Valade2022} independently improved on the method presented in \citet{Graziani2019} by replacing the Gibbs-sampling with an HMC technique and taking advantage of the modern GPU-compatible python computation library \textsc{Jax} \citep{Bradbury2018}, accelerating the numerical exploration by orders of magnitude\footnote{We note that the original implementation of \citet{Valade2022} is based on \textsc{TensorFlow} \citep{Dillon2017} and has since been ported to \textsc{Jax} \citep{Bradbury2018}.}. An application of the {\sc HAMLET} methodology to the Cosmicflows-4 catalog \citet[CF4]{Tully2023} is presented in \citet{Valade2024} while \citet{Boruah2020} discusses the application of their method to a composite dataset. We note that a third separate single-CPU implementation of the HMC version of \citet{Graziani2019} has been developed and applied to Cosmicflows-4 by \citet{Courtois2023}. 

The peculiar velocity based reconstruction methods discussed above are based on the linear relationship between density and velocity and are thus practically limited to where peculiar velocities are accessible: the late-time Universe. The first fully Bayesian reconstruction of initial conditions from velocity data was proposed by \citet{Prideaux-Ghee2022}. Here, the dynamics of particles are modeled with the Zeldovich approximation \citep{Zeldovich1970} and the late-time velocity field is computed from the reconstructed Cloud in Cell density field \citep[CiC;][]{Birdsall1969} within the linear theory. This approach was further improved on by \citet{Bayer2023} by replacing the Zeldovich approximation with a particle-mesh gravity solver \citep[\textsc{FlowPM};][]{Modi2021} and by avoiding the use of linear theory in the late-time Universe.

In this work, we present \HamletPM{}, an extension of the \Hamlet{} method \citep{Valade2022} that enables constraining the initial conditions directly from $z=0$ velocity data. The linear model of the former is replaced with a differentiable particle-mesh gravity solver \citep[\pmwd{};][]{Li2022}. To tackle the issue of empty cells, the $z=0$ velocity field is constructed using a hybrid approach mixing a CiC technique in dense regions and linear theory. The initial conditions are then augmented with random small-scale power and run to $z=0$ with the \textsc{Arepo} code \citep{Springel2010} resulting in 100 simulations of the Local Universe with 512$^3$ particles in a box of size $500\Mpch$ constrained up to $z\sim0.05$ ($160\,h^{-1}{\rm Mpc}$).

Throughout this paper we use $\Omega_m = 0.3$, $\Omega_b=0.0471$ and $H_0 = 74.6 \kms/\Mpc$. Consistent with other local measurements, the high value of $H_0$ is fitted to the Cosmicflows-4 data so as to ensure no global radial in- or out- flow \citep{Tully2023,Duangchan2025}.

\section{Data}
\label{sec:data}

Peculiar velocities catalogs are built by complementing redshift surveys with physical distance measurements, so as to isolate the radial component of each galaxy's peculiar velocity $(v_r)$ from its Hubble expansion \citep{Hubble1929}. To first order we write:
\begin{equation}
    v_r = cz - H_0 d
    \label{eq:vr_obs}
\end{equation}
where $H_0$ is the Hubble constant and $c$ the velocity of light. $z$ and $d$ are observed measurements of redshift and physical distance. Several distance estimators exist, with relative uncertainties ranging from $5\%$ to $20\%$, depending on the method. Propagated to the peculiar velocity in \cref{eq:vr_obs}, an uncertainty on the distance estimator of $5\%$ (\resp{} 20\%) results in a signal to noise ratio of less than unity, for galaxies beyond $80\Mpc$ (\resp{} $20\Mpc$). Namely at these distances the peculiar velocity error is greater than the expected value of the peculiar velocity itself.

In this work, we employ the grouped version of the Cosmicflows-4 \citep[CF4;][]{Tully2023}, the largest compilation of peculiar velocities to date, with about 56\,000 galaxies gathered in more than 38\,000 groups\footnote{Note that we use the term ``group'' generically, even for groups composed of a single galaxy.}. The grouping offers two advantages over the raw data, as it mitigates the Finger-of-God effect in massive galaxy clusters \citep{Kaiser1987} and improves the quality of the constraints by averaging distance measurements.  All in all, the sample is rather isotropic up to $z\approx 0.055$ but reaches to $z = 0.1$ in the SDSS footprint \citep{York2000}.

Rather than in a single survey, CF4 is a composite catalog. The majority of its distances are derived from the Fundamental-Plane scaling relationship \citep[FP;][]{Djorgovski1987}. The SDSS peculiar velocity (SDSS-PPV) catalog contains 37\,000 distances \citep{Howlett2017} while 6dFGRSv \citep{Campbell2014} contributes 8\,800 distances. The former is limited to the SDSS footprint in the northern hemisphere and reaches $z=0.1$ while the latter is limited to the southern sky and reaches up to $z=0.055$. CF4 contains more than 12\,000 distances obtained with the Tully-Fisher (TF) scaling relationship \citep[TF;][]{Tully1977} from \citet[9\,000 distances]{Kourkchi2022} and other catalogs. The TF sample is relatively isotropic and contained within $z=0.055$ with a peak around $z=0.02$. About 1\,000 type 1a Super Novae (SN1a) from various sources, notably SH0ES \citep{Riess2022} and PantheonPlus \citep{Brout2022} have been added to CF4. They provide whole sky distances up to $z=0.1$, peaking at $z=0.02$.  Finally, high-quality estimators such as Cepheids \citep{Leavitt1912}, Tip of the Red Giant Branch \citep[TRGB;][]{Frogel1983} or Surface Brightness Fluctuation \citep{Tonry1988} are found up to $z=0.02$. 

While the errors of the redshift are small (about $50\kms$ for all groups) and normally distributed, particular attention must be paid to the distance measurements. At the first order, the distances $(d)$ are derived from distance moduli $(\mu)$
\begin{equation}
    d = 10^{\mu / 5 - 5} \Mpc
    \label{eq:d_to_mu}
\end{equation}
Distance moduli measurements are assumed to have normally distributed errors $\mathcal{N}(\mu, \sigma_\mu)$ with an amplitude $\sigma_\mu$ depending on the measurement method, ranging from about $0.05 \mg$ for the best measured groups to $0.5\mg$ for the large majority of the entries. Uncertainties on distance modulus translate into relative errors on distances ranging from $5\%$ to $20\%$, leaving a very low median signal-to-noise ratio of less than 10\%. Furthermore, \cref{eq:d_to_mu} transforms the normally distributed uncertainty on the distance modulus into log-normally distributed errors on the distance measurement. If the data are naively interpreted, this leads to a spurious infall \citep[see][]{Hoffman2021}. Last but not least, a typical uncertainty of order $0.1\mg$ remains in the inter-calibration of the different subcatalogs of CF4. This work does not address this issue, whose influence on the large-scale flow may be significant \cite[fig. 10]{Whitford2023}.

In this work, we limit the CF4 data to a redshift velocity of $cz_{\max} = 16,500\kms$ \citep{Magoulas2016}. This value is consistent with the boundary of 6dF and removes only a few points from the TF galaxy sample, but severely chops the SDSS PV data, thus removing the  highly lop-sided shape of the CF4 catalog. This choice is motivated by technical limitations of the method described below, which allows us to focus our computing power towards the core of the sample, where the errors are smaller and the data more isotropic.

\section{Method}
\label{sec:method}

This work builds on the previously presented \Hamlet{} code, whose details can be found in \citet{Valade2022}. A quick summary is given below, followed by a thorough explanation of how modifications have been made to produce initial conditions.

\subsection{Bayesian framework}

The proposed method is written in a Bayesian context. The aim is to derive and estimate the \emph{posterior} probability distribution of a set of free parameters given some fixed observations. In our case the free parameters are the distances of the constraints $\lD = \{d_i\}$ for $i \leq n_{\rm obs}$ and the Fourier modes of the high redshift over-density field projected on a grid $\Delta = \{\hat{\delta}_{\vk}\}$ associated with the accessible modes $\{\vk_j\}$ for $j < n_{\rm modes}$. We denote the observations\footnote{The superscript ``obs'' refers to observational data so as to avoid confusion with the similar quantities computed by the model.} by their distance moduli $\lM = \{\muo_i\}$ and their redshifts $\lZ=\{\zo_i\}$. Note that distance moduli are in fact not direct observations, though we consider them as such in this model for the sake of simplicity. A better modeling, such as that of \citet{Stiskalek2025}, may be explored in further works. 

Following Bayes theorem, the posterior probability can be written
\begin{equation}
    P(\Delta, \lD \mid \lM, \lZ) \propto P(\lM, \lZ \mid \Delta, \lD) P(\Delta, \lD) 
    \label{eq:posterior}
\end{equation}
where the denominator $P(\lM, \lZ)$ has been omitted as it is not a function of the free parameters. The two terms on the right-hand side are respectively named the \emph{likelihood} and the \emph{prior} distribution. Implicit to these equations is the underlying mathematical model linking the free parameters and the observations. \Cref{sec:obs_model} details the description of the observations and \cref{sec:field_model} focuses on the modeling of the density and velocity fields. 

\subsection{Observational model}
\label{sec:obs_model}

The mathematical derivation of the model is detailed here, while a visual summary is presented in \cref{fig:dag} under the form of a Directed Acyclic Graph (DAG).

\subsubsection{Likelihood}

The forward model is composed of two parts. First, the observed quantities $(\mu, z)$ need to be computed from the model parameters $(d, \Delta)$ for each constraint, alongside with a few intermediaries:
\begin{align}
    &v_r = \vv(d \hr) \cdot \hr, \\
    \label{eq:zc_from_d}
    &z_c = \frac{2}{3 \Omega_m} \left(1 - \sqrt{\frac{3 \Omega_m H_0}{c} d} \right), \\ 
    &z = (1 + z_c) (1 + v_r / c) - 1, \\
    \label{eq:dl_from_d}
    &d_L = (1 + z + (\vv_{\rm CMB} \cdot \hr) / c) d, \\
    \label{eq:mu_from_dl}
    &\mu = 5 \log_{10}(d_L) + 25
\end{align}
where $\vv$ is the 3D peculiar velocity field, $v_r$ is the radial peculiar velocity, $\hr$ is the unit direction vector, $z_c$ and $z$ are respectively the cosmological and full redshifts, $d_L$ the luminosity distance, $\vv_{\rm CMB}$ is the 3D velocity of the Milky Way with respect to the CMB and $\mu$ the distance modulus. The form of \cref{eq:dl_from_d} is given by \citet{Calcino2017}.

Once the observables have been derived, their probabilities can be written. The likelihood $P(\lM, \lZ \mid \Delta, \lD)$ reads
\begin{equation}
    P(\lM, \lZ \mid \Delta, \lD) = \prod_i \lN\left[\muo_i, \sigma_{\mu, i}\right](\mu_i) \times \lN\left[\zo_i, \sigma_z\right](z_i)
    \label{eq:likelihood}
\end{equation}
where $\lN[m, s](x)$ denotes the gaussian distribution with a mean $m$ and a scatter $s$ evaluated at $x$. Note that \cref{eq:likelihood} neglects the error coming from the inter-calibration process. 

\subsubsection{Selection function and prior distribution}

Observational constraints (\eg{} flux-limitation) restrict the number of observed objects. While a proper description of these effects is important for an unbiased Bayesian analysis, such as that detailed in \citet{Kelly2008}, the composite nature of the Cosmicflows-4 catalog makes the modeling of the selection function an arduous task. We then opt for a simplified approach, detailed in \cref{app:model}, by considering an ad-hoc prior function 
\begin{equation}
    P(\lD) = \prod_i \hat{\lH}_{d_z}(d_i)
\end{equation}
where $\hat{\lH}_{d_z}$ is the distribution of observed redshift distances computed from the observed redshifts, smoothed with a $3\Mpch$ gaussian kernel. Note that alternative approximations exist, such as \citet{Stiskalek2025}.

\subsection{Velocity field model}
\label{sec:field_model}


To reconstruct the late time velocity field, we employ an n-body gravity particle-mesh \citep[PM;][]{Kates1991} solver. Dubbed \pmwd{}, the implementation used is written in JAX, GPU-accelerated, differentiable and can be readily plugged into our \Hamlet{} implementation \citep{Li2022}. These are gravity only simulations and thus do not include any baryonic effects. To keep the inference loop tractable, we limit ourselves to $128^3$ dark matter particles in a box of $500^3(\Mpch)^3$, resulting in a mass resolution of $4.9\times 10^{12} \Msh$. To resolve the gravitational potential, particles are gridded on a $128^3$ mesh, leading to a force resolution of $3.9\Mpch$. 

To enforce the power spectrum, we set the following prior on the Fourier modes of the initial density field:
\begin{equation}
    P(\Delta) = \prod_k \frac{-|\hat{\delta}_{\vk}|^2}{\lP(|\vk|)} 
    \label{eq:prior}
\end{equation}
where $\lP$ is the matter-matter power-spectrum. This power spectrum is generated within \pmwd{} using an amplitude $A_s = 2\times10^9$ and a spectral number $n_s=0.96$. The ICs are then brought to a redshift $z=z_{\rm PM}$ with a second order Lagrangian Perturbation Theory \citep[2-LPT;][]{Buchert1993} and further evolved to $z=0$ with $n_{\rm PM}$ PM steps. 

A first order Cloud in Cell \citep[CiC][]{Birdsall1969} algorithm is used to obtain a field representation of the density field $(\rhocic)$ and velocity field $(\vcic)$ at redshift $z=0$. We set the CiC grid resolution to that of the particles' grid, \ie{} a cell size of $3.9\Mpch$. 

Constructing the velocity field from an $N$-body simulation may lead to issues in both dense and  empty  regions \citep{Hahn2015}. Notably, the CiC of the velocity cannot be correctly estimated in void regions, where empty cells are found. To rectify this issue, we employ a linear model of the velocity field in nearly empty regions. Linear theory is accepted to be accurate in the early universe or at low redshift on large scales (above $5-10\Mpch$). The linear relationship between density and velocity can still be applied in an evolved universe as a simplifying approximation, to derive the velocity field from the non-linear over-density field  
\begin{equation}
   - H_0 f \deltacic = \vnab \cdot \vlin
    \label{eq:linear_dyn}
\end{equation}
where $f$ is the linear growth factor and $\delta=(\rho-\bar{\rho})/\bar{\rho}$ is the density contrast. This approximation is also made by \citet{Prideaux-Ghee2022}. \Cref{eq:linear_dyn} tends to remain accurate at all scales in void regions and is well defined even in empty cells, however, it does overestimate the velocity in dense regions, leading to spurious dynamics in and around dense structures. 

The velocity field used for the inference is thus a hybrid of these two fields, using the CiC representation in populated cells and the linear representation in emptier regions: 
\begin{align}
    &\vv = w\big(\rhocic\big) \vcic + (1 - w\big(\rhocic\big))\vlin \\
    &w\big(\rhocic\big) = \frac{1}{2} \left(1 + {\rm erf}\left(\frac{\log_{10}\big(\rhocic\big) - \alpha}{\beta} \right)\right).
\end{align}
where $\rho^{\rm CiC}$ is the over density in a cell. Note that when an empty cell $\rho^{\rm CiC}=0$ gives $w\big(\rho)\big)=0$. 
In this work, we use $z_{\rm PM} = 19$, $n_{\rm PM} = 5$, $\alpha = -0.3$ and $\beta = 0.1$. These values are based on preliminary tests and shall be more deeply discussed in further studies.

\subsection{Running the constrained simulations}

The posterior function described above is constructed with our \texttt{banana} framework\footnote{\url{https://github.com/anvvalade/banana}} and explored with a Hamiltonian Monte Carlo algorithm \citep[HMC;][]{Neal2011} implemented by the python library \texttt{blackjax}\footnote{\url{https://github.com/blackjax-devs/blackjax}}. This exploration is complicated by  non-linearities, necessitating significantly more computational power than the linear model presented in \citet{Valade2022, Valade2024}. To reduce self-correlation in the random sample, ten chains of 5\,000 steps are run. For each chain, only one sample every 500 steps is considered, leading to a total of $10 \times 10 = 100$ constrained initial conditions. We verified that the internal correlation on each Monte Carlo chain was minimal (see \cref{app:mc_conv}). Each Monte Carlo chain took about 10 hours to run on an NVIDIA-V100 GPU, meaning we were able to create statistically independent ICs every hour of computation. 
  
The initial conditions inferred from the data according to the method described above, have a relatively low resolution of $128^3$ cells. To construct higher resolution initial conditions, random small scale power is injected so as to reach a resolution of $512^3$ particles. For this simple purpose, we have developed our own routine. The resulting particle mass is $7.8\times10^{10}\Msh$. The initial conditions are brought to $z=63$ with the 2-LPT and then evolved to $z=0$ with the AREPO gravity solver \citep{Springel2010}. Dark matter halos are produced on the fly in AREPO according to a Friend-of-Friend algorithm, using the standard $l = 0.2$ linking length. Only halos with more than 32 particles are considered, leading to a minimum halo mass of $2.5\times10^{12}\Msh$.

\section{Quality assessment of the constrained simulations}

In the last few years,  assessing the quality of such simulations has moved from a qualitative ``visual'' approach to a more quantitative one. So far, two metrics have emerged that compare constrained simulations to observations, namely the Local Universe Model (LUM, \cta{Pfeifer2023}) focusing on galaxy clusters and the Velocity Field Olympics \citep[VFO;][]{Stiskalek2025} studying the velocity field. Going even further, \cta{McAlpine2025} suggested a three-level verification program, from essential to more ambitious, that we summarize as follows: (1) verifying that the simulation is compatible with the underlying cosmological model; (2) quantifying the quality of the constrained volume via rigorous metrics and (3) reproducing more observables \citep[\eg{} tSZ;][]{Stiskalek2026a}, \eg{} through hydro-dynamical re-simulations \citep[\eg{}][]{Hernandez-Martinez2024}. 

Joining the movement toward standardization, this manuscript endeavors to fulfill tiers (1) \& (2), leaving however tier (3) for later studies. 

\subsection{Comparison to the literature}

The advent of quantitative metrics enables a fair comparison between constrained simulation suites, as well as ``direct'' reconstructions of the Local Universe. Here, we give a quick overview of the methods appearing in the result and discussion sections. 

Two sets of velocity-based constrained simulations are discussed. First are the nine CLONES simulations \citep{Sorce2018} alongside with an accompanying high-resolution baryonic version of one of the CLONES simulations \citep{Dolag2023, Hernandez-Martinez2024} dubbed SLOW. Second, the one-hundred constrained simulations of \cta{Pfeifer2023} are considered. Both of them apply the CLUES pipeline (WF/RZA/CRs) but they differ in their debiasing methodologies, presented respectively in \citep{Sorce2018} and \citet{Hoffman2021}; as well as in the data sources, Cosmicflows-2 \citep{Tully2013} for the former and Cosmicflows-3 for the latter \citep{Tully2016}. The direct reconstructions of the Local Universe of \citet{Courtois2023} and \citet{Valade2024} are also discussed, both of which apply a forward-modeling method to the Cosmicflows-4 dataset.  

We also consider two sets of redshift-based constrained simulations, namely CSiBORG1 \citep{Bartlett2021} and Manticore-Local \citep{McAlpine2025}. Both simulation suites stem from the Aquila consortium and apply a non-linear forward-modeling to augmented versions of the 2MASS catalog \citep{Huchra2012}. The core difference between CSiBORG1 and Manticore-Local is the replacement of a particle-mesh gravity solver by COLA \citep{Tassev2013}, as discussed in \citet{Stopyra2024}, alongside with the addition of new constraints to insure that the initial conditions remain normally distributed.  Lastly, two direct reconstruction methods are discussed, namely \citet{Carrick2015} and \citet{Lilow2024}, both working with the 2MASS catalog of redshifts. The former is a relatively naive estimation of the density field from direct galaxy count, while the latter utilizes neural networks trained on full $n$-body simulations.

\subsection{Recovering galaxy clusters}

Massive galaxy clusters are found at the nodes of the large-scale structure. Such high mass objects are relatively rare. Their abundance, properties and environment in a given volume are thus subject to cosmic variance. Recovering them in a controlled environment with constrained simulations thus holds a significant scientific advantage.

\subsubsection{The Local Universe Model (LUM)}

The Local Universe Model (LUM) is a framework proposed by \cta{Pfeifer2023} which aims to quantify the deviation from randomness of a constrained simulation, namely how successful a given attempt is. In essence, it looks at massive simulated haloes in the neighborhood of the position of an observed cluster and quantifies the probability of the null-hypothesis: \emph{``this halo stems from a random simulation''}.  
In practice, a $p$-value depending on the separation to the target cluster and the mass of the halo is computed for each halo $M \geq 10^{14} \Msh$. For example, a halo of mass $5\cdot10^{14}\Msh$ at a distance of $10\Mpch$ from the target cluster has a value $p = 0.01$: one needs to run (on average) 100 (random) simulations to find a halo (at least) that massive and at least that close to the target cluster. Details on our implementation of the LUM are found in \cref{app:lum}.

\subsubsection{Optimizing the Local Universe Model (opt-LUM)}

While introducing a relevant metric, the LUM overlooks several aspects of the problem it aims to solve. Indeed, it is unable to disentangle \emph{systematic} positional errors in the halos distribution from random scatter. This systematic shift may be a shortcoming of the method or may stem from an observational effect, such as uncertainties on the observed cluster position, whether this uncertainty stems from redshift space distortion or direct distance estimations. Suppose that for a given cluster, a suite of constrained simulations produces a cloud of massive, halos whose center is slightly shifted with respect to the observed cluster. In this case the LUM would penalize the offset, thereby returning high $p$-values. We introduce here the opt-LUM metric, which is designed to capture such a tight gathering of massive halos -- provided there is one in the neighborhood of the observed cluster -- while disregarding less massive and less densely-scattered halos.

The opt-LUM is a simple algorithmic amendment to the LUM. In essence, for each cluster, the method iteratively shifts the target position until it is at the center of mass of the matched halos. The procedure unfolds as follows. At the initial time, the targets are set at the observed clusters positions (since we are looking for matched haloes in the vicinity of a given observed cluster. Then, (1) halos are matched with the targets using the LUM approach described above and (2) for each cluster, the new target position is constructed from the matched halos as the ``center of mass'' weighted by the simulated halo's mass. If each realization is indicated by the subscript ($i$), then the position of the target, at a given iteration is 
\begin{equation}
  \vr_{\rm target} = \frac{\sum_i \vr_i w_i}{\sum_i w_i}, \quad w_i = \log_{10}\left(\frac{M_i}{\min M_i}\right).
  \label{eq:optlum}
\end{equation}
Steps (1) and (2) are repeated until 
\begin{equation}
  \max \left(p_{\rm val}^{(s + 1)} - p_{\rm val}^{(s)} \right) < 10^{-4}
\end{equation}
where $p_{\rm val}$ is the p-value, $(s)$ denotes the $s$-th iteration step and the maximum is taken over all considered clusters. The above threshold is reached in less than 10 iterations. 

Essentially, the opt-LUM gives higher detection rates than those of the pure LUM, yet both values cannot be directly compared. The \pval{}s yielded by the opt-LUM metric are with respect to the center of mass of the cloud of halos, while the \pval{}s of the LUM are with respect to the observational cluster. The opt-LUM is not meant to replace the LUM metric, but rather to augment it.

\subsubsection{Selected clusters in the Local Universe}
\begin{table}
  \caption{Selected clusters to which the Local Universe Model is applied. Positions are derived from the MCXCII X-Ray catalog \citep{Sadibekova2024}. The redshift distances have been transformed into comoving using our cosmology.}
  \label{tab:clusters}
    \begin{center}
        \tiny
    \begin{tabular}{l|ccc|r}
      Cluster       & SGX                            & SGY    & SGZ    & MCXII name           \\
      ~             & \multicolumn{3}{c|}{[Mpc/$h$]}                    & ~                   \\
      \hline                                                                                
      Virgo         & -2.68                          & 11.70  & -0.48  & Virgo                \\
      Fornax        & -1.47                          & -11.24 & -10.23 & Fornax               \\
      Centaurus     & -30.63                         & 13.35  & -6.76  & A3526                \\
      Hydra         & -23.91                         & 20.52  & -24.23 & A1060                \\
      Norma         & -46.10                         & -6.49  & 5.71   & A3627                \\
      Perseus       & 50.65                          & -10.88 & -13.16 & A426                 \\
      Leo           & -2.50                          & 62.72  & -11.65 & A1367                \\
      Coma          & 0.41                           & 68.17  & 9.94   & Coma                 \\
      Ophiuchus     & -59.99                         & 7.18   & 57.50  & Ophiuchus            \\
      Hercules      & -38.75                         & 73.81  & 64.77  & A2063                \\
      Shapley (A)   & -102.02                        & 55.41  & 4.26   & A3571                \\
      Shapley (B)   & -121.81                        & 73.54  & -3.40  & A3558                \\
      \end{tabular}
    \end{center}
\end{table}

\begin{figure*}
    \begin{center}
        \includegraphics[width=.49\textwidth]{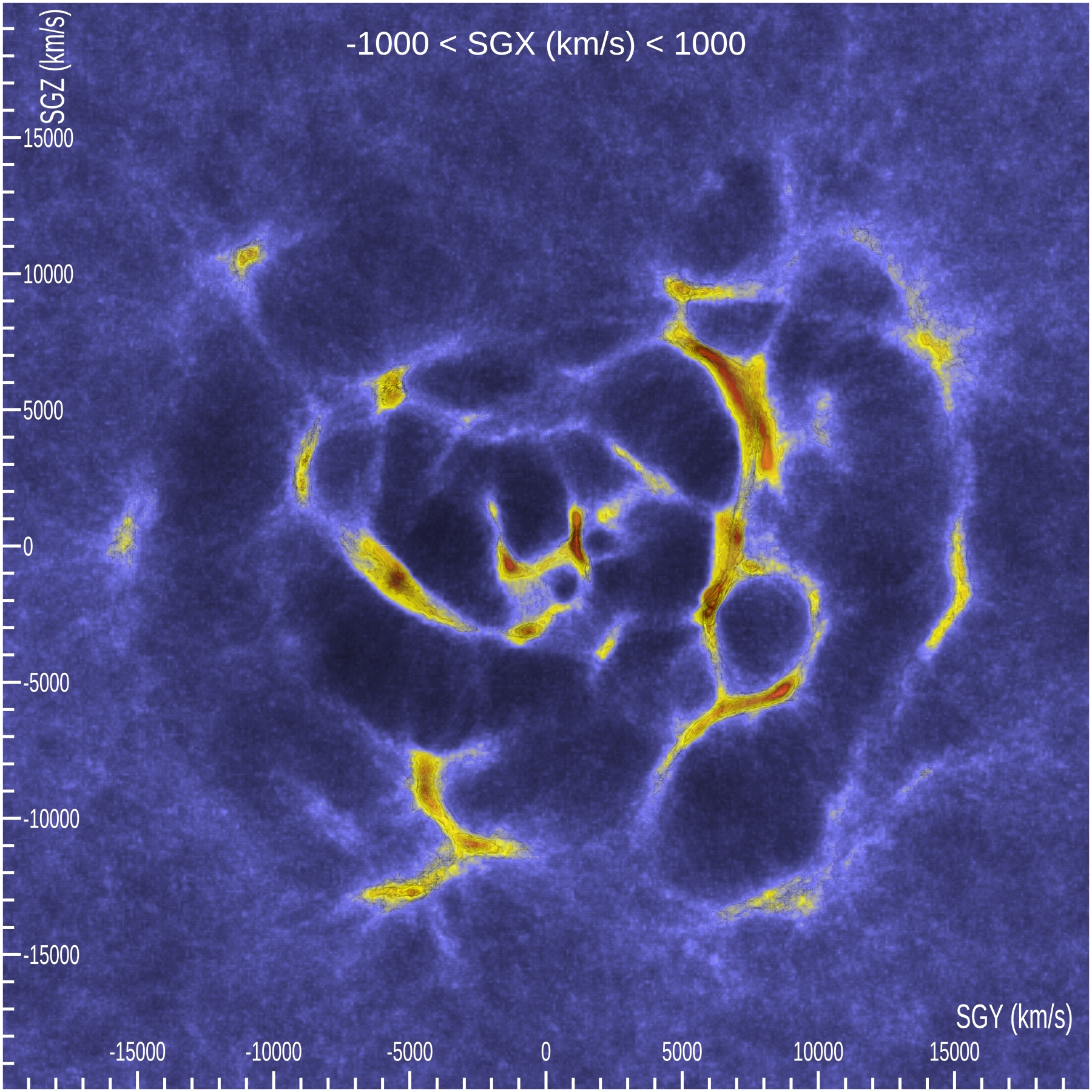}  
        \includegraphics[width=.49\textwidth]{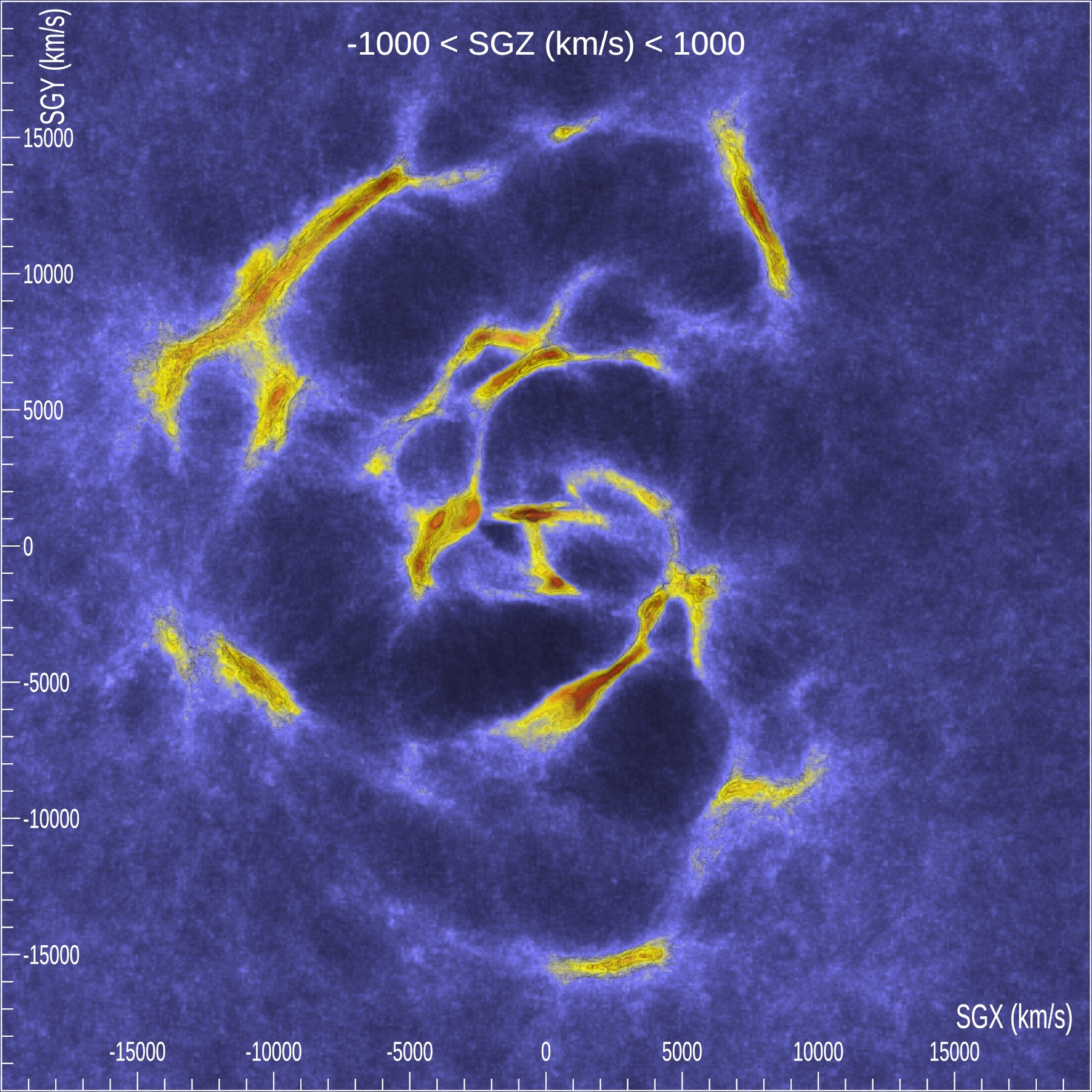}  
    \end{center}
    \caption{Slices of the mean density field through the super-galactic planes SGX=0 and SGZ=0, cropped to the constrained region. The mean density is computed as a geometrical mean, \ie{} $\tilde{\rho} = \left(\prod_i \rho_i\right)^{1/n}$. This approach increases the readibility and enables comparison with \citet{Hoffman2018,Pfeifer2023}.  }
    \label{fig:xz-slices}
\end{figure*}

\begin{figure}
    \begin{center}
        \includegraphics[width=0.95\columnwidth]{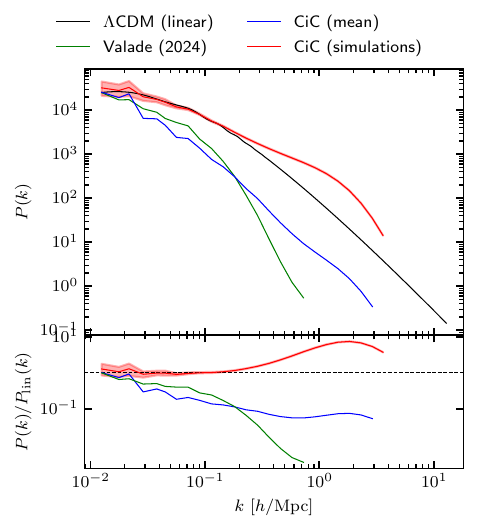}
    \end{center}
    \caption{Power spectrum of the initial as well as the evolved density fields compared with the linear power spectrum. For the reconstructed fields, the solid lines and the shades  respectively represent the median and $68\%$ interval around it over the 100 constrained simulations. The initial (\resp{} evolved) density fields have a resolution of $128^3$ (\resp{} $512^3$) nodes. }
    \label{fig:pk}
\end{figure}

\begin{figure}
    \begin{center}
        \includegraphics[width=0.95\columnwidth]{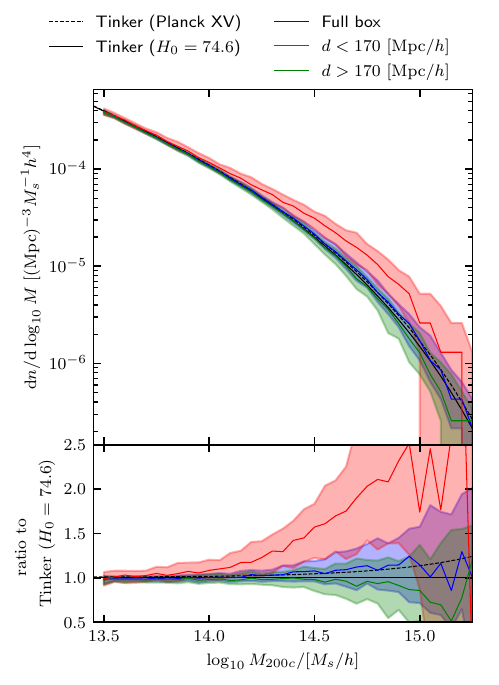}
    \end{center}
    \caption{Halo mass function of the entire volume, in the constrained volume and in the remaining, unconstrained volume. The shades indicate the 1$\sigma$ confidence level derived from the 100 constrained simulations. The empirical halo mass function of \citet{Tinker2008} is shown twice; once with the cosmology of the simulation suite and once with the standard Planck 15 values \citep{PlanckCollaboration2016}. The lower panel shows the ratio to \citet{Tinker2008} with the cosmology of the simulations suite.}
    \label{fig:hmf}
\end{figure}

\begin{figure}
    \begin{center}
        \includegraphics[width=\columnwidth]{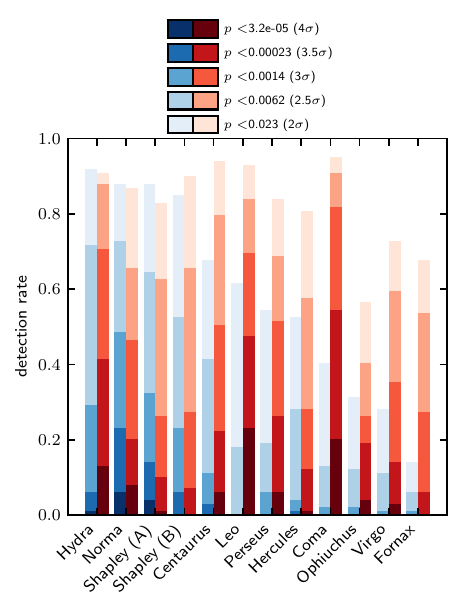}
    \end{center}
    \caption{Detection rate of major (or nearby) clusters of the Local Universe, defined in \cta{Pfeifer2023} as the rate of simulations meeting the upper p-value threshold for each given cluster. Results for the LUM approach are given in blue and results for the opt-LUM algorithm (see \cref{eq:optlum}) in red.}
    \label{fig:drate}
\end{figure}

\begin{figure*}
    \begin{center}
        \includegraphics[width=\textwidth]{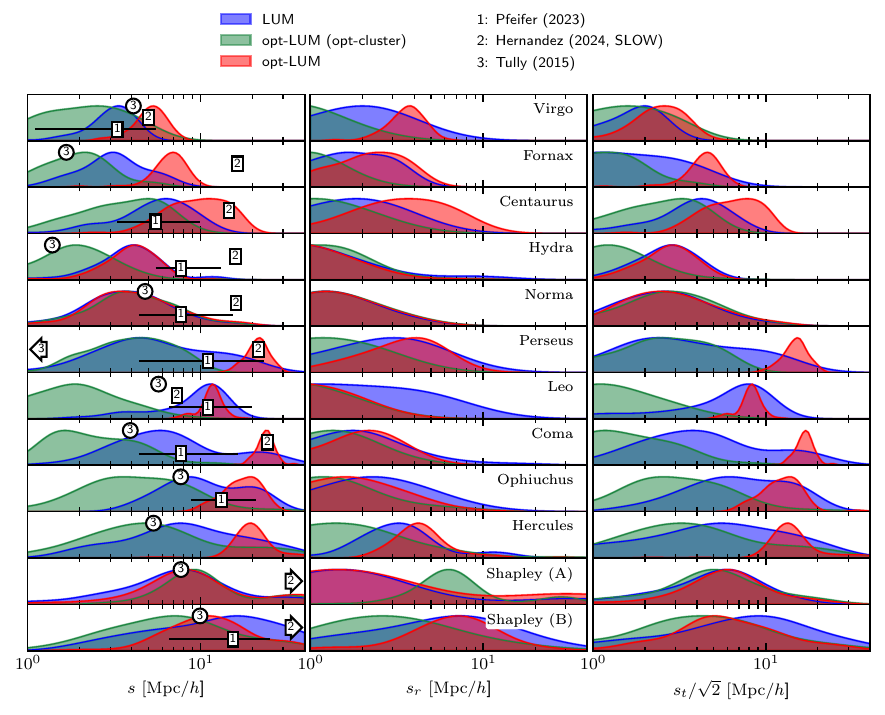}
    \end{center}
    \caption{Posterior separation of the matched halos to the observational cluster. The left, central and right columns respectively represent the 3D, radial and ortho-radial separations. The blue and red curves respectively correspond to halos matched with LUM and opt-LUM methods. Values in green indicate the separation between the halos matched by the opt-LUM method and the associated shifted target. The ortho-radial component is normalized by a factor $\sqrt{2}$ so as to enable comparison with the radial component (for an isotropic scatter, $|s_t| / |s_r| = \sqrt{2}$). Clusters are sorted by increasing distance. The separations presented in \cta{McAlpine2025}, all below $1\Mpch$, are not plotted here.}
    \label{fig:sep}
\end{figure*}

\begin{figure*}
    \begin{center}
        \includegraphics[width=.9\textwidth]{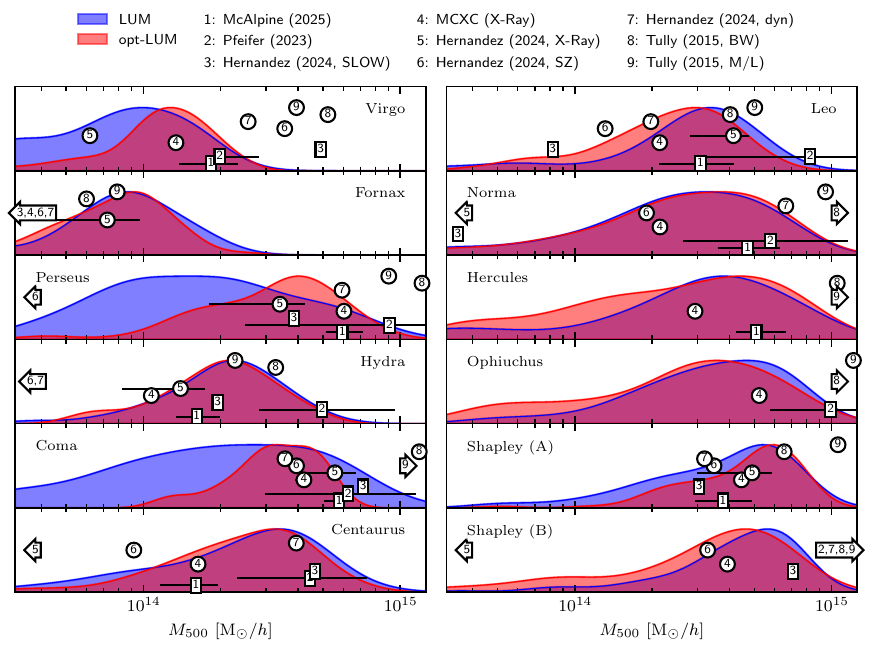}
    \end{center}
    \caption{Posterior mass distributions of major (or nearby) clusters of the Local Universe. To facilitate comparison with mass estimates in the literature, the $M_{500c}$ is used, that is, the mass in the inner region of the halo such that $\rho > 500 \rho_c$. Values of \cta{Pfeifer2023}, provided in $M_{200c}$ have been divided by an empirical factor $3/2$. Values from the literature in squares are taken from constrained simulations, while circled values are observational. Arrows pointing to the left (\resp{} right) indicate that the value is out of the plotted range. Clusters are sorted by increasing posterior mean mass.}
    \label{fig:mass}
\end{figure*}

\begin{figure}
    \begin{center}
        \includegraphics[width=\columnwidth]{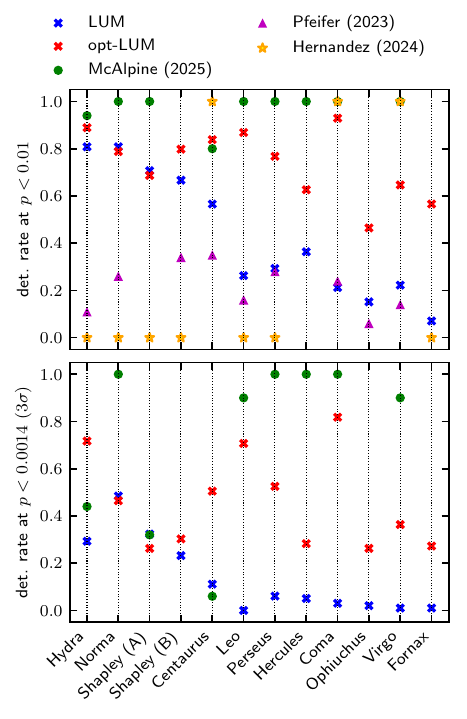}
    \end{center}
    \caption{Comparison of the detection rates between several simulation suites. For compatibility with \cta{Pfeifer2023} and \cta{McAlpine2025}, the upper panel gives the detection rate at $p=0.01$, while the lower panel shows the detection rate at $p = 0.00014$ to improve comparison and avoid saturated values at $p=0.01$. To report results from \cta{Hernandez-Martinez2024}, we set a value of one if the cluster is detected with at least a \pval{} lower than the threshold, zero otherwise. This is however not a proper detection rate.}
    \label{fig:drate_cmp_rss}
\end{figure}

\begin{figure}
    \begin{center}
        \includegraphics[width=.9\columnwidth]{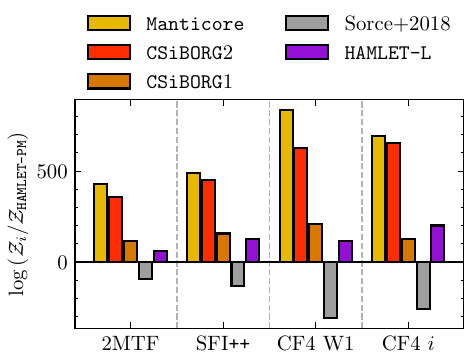}
    \end{center}
    \caption{Log of the evidence of several constrained simulation suites as well as the linear reconstruction \citet[\Hamlet-L;][]{Valade2024} with respect to this work. Manticore-Local \citep{McAlpine2025}, CSiBORG2 \citep{Stopyra2024, Stiskalek2025} and CSiBORG1 \citep{Bartlett2021} are constrained with redshift data while CLONES \citep{Sorce2018} is constructed on velocity data \citep[CF2; ][]{Tully2013}.}
    \label{fig:vfo_wrt_hamlet}
\end{figure}
 
There is so far no consensus in the constrained simulation community on a benchmark catalog of clusters to be reproduced, in order to validate one's bid at a constrained simulation. Indeed, the building of such a wishlist depends on the data used -- mostly sky coverage and depth -- but also on the constraining power of both data and reconstruction method, as less massive halos are harder to recover than more massive ones. The choice of studied clusters seems then to have grown organically in each group. 

In this work, twelve clusters are studied. These have been selected from the largest clusters, so as to probe different well reconstructed regions while maximizing the overlap with the pre-existing literature. The positions of the clusters have been extracted from the MCXC-II catalog of X-ray clusters \citep{Sadibekova2024}. Estimations of mass have been extracted from MCXC-II; the literature review from \cta{Hernandez-Martinez2024} (Table A2) which includes mass estimations derived from X-Ray, Sunyaev-Zeldovich and dynamical probes; and finally mass measurements from \citet{Tully2015} comprising values from an empirical luminosity-mass relationship and from dynamical probes.

\subsection{Assessing the velocity field}

The assessment described above focuses on measuring how well a constrained simulation reproduces the local cluster positions -- namely the main sources of gravitational attraction. \citet{Stiskalek2025} offers the first quantitative benchmark of constrained simulations based on the reconstructed velocity field, known as the Velocity Fields Olympics (VFO).

In this framework, the reconstructed velocity fields are compared to observational catalogs of peculiar velocities. Accordingly, \citet{Stiskalek2025} considers primarily the Tully-Fischer datasets SFI++, 2MTF, CF4-i and CF4-W1 \citep[\resp{}][]{Masters2006,Masters2008,Kourkchi2020a,Kourkchi2020b}, while the analysis is extended to other catalogs in the appendix. For each set of data, an elaborate forward model is built so as to reproduce the linewidth, as well as biases \citep[\eg{} in-homogeneous Malmquist bias;][]{Strauss1995}. In absence of absolute calibration, the outcome is independent of the assumed $H_0$ constant. Yet, it does depend on smoothing of the reconstructed fields and on the overlap between the data used to constrain the simulations and the test data. 

The results of the VFO consist of a Bayesian evidence ($\lZ$) for each test data set, which quantifies the ability of a given model to match a data set. The higher the evidence, the more compatible the reconstructed field is with the test data. Note that the evidence scales with the size of the considered data. The approach also fits for the global parameter $\beta$, which we interpret in this work as an amplification parameter of the velocity field with respect to the data. A $\beta$ factor superior (\resp{} inferior) to unity means that the tested method under-estimates (\resp{} over-estimates) the contrast of the observed velocities.

\section{Results}
\label{sec:res}

\subsection{Density field}

\label{sec:res:lcdm}

To gain insight on the constraining level of the method, \cref{fig:xz-slices} show two slices through the density field averaged over the 100 constrained simulations. Outside the constrained volume, the density field converges to the cosmic average. This is the expected behavior within the adopted Bayesian framework and seems to indicate that the Monte Carlo chains are converged, although it cannot be excluded that the convergence inside and outside the constrained volume are different. 

On the other hand, structures within the constrained volume are persistent, meaning they are ``constrained''. Prominent voids are visible as well as a system of filaments -- or cut walls -- linking denser nodes, as expected from observations of the large-scale structure \citep{Bond1996,Libeskind2018}. Non-linear structures appear much more pronounced than in linear reconstructions of the Local Universe \citep[\eg{}][]{Graziani2019, Sorce2023, Courtois2023, Hoffman2024, Valade2024} or previously published mean density fields of velocity-based constrained simulations \citep{Hoffman2018, Pfeifer2023}, although we note that this might be partially due to smoothing. 

As in previous forward-modeling reconstructions from radial peculiar velocities \citet{Graziani2019, Valade2024}, transverse structures appear much better captured by the model and the data than radial structures, leading to a ``bulls-eye'' visual impression.
\Cref{fig:pk} presents the power spectra of the 100 constrained simulations.  The power spectrum of the initial conditions appears to match well the \LCDM{} prediction, while we note that the power appears to be slightly too high over all the probed range which is discussed alongside with other results in \cref{sec:dis}. The evolved power spectra follows the \LCDM{} linear power spectrum up to $k \approx 0.2\hMpc$, frequency beyond which the evolved power spectra depart from linear predictions due non-linearities in the field. 

The halo mass function is shown in \cref{fig:hmf} for the full box, the constrained region ($d < 165 \Mpch$) and in the remaining volume. Two reference halo mass functions are given, one for the ``standard'' Planck-XV cosmology and one for the cosmology used in this work, both computed with the empirical model of \citet{Tinker2008}. The current cosmology is expected to produce less massive clusters than the standard Planck-XV cosmology, reaching about 20\% under-abundance of $10^{15}\Msh$ halos. The halo mass distribution in the full box is consistent with \LCDM{}. The scatter between realizations includes the expected distributions of both Planck-XV and custom cosmologies. Yet, a marked over-abundance of massive clusters is found in the constrained regions: starting at  $10^{14}\Msh$ we produce $\sim$ 5\% more haloes than  expected whereas at $10^{15}\Msh$ this increases to roughly double the expected number of haloes. This over-abundance is in strong tension with the underlying cosmological model and at odds with observations of the Local Universe \citep{Bohringer2020,Dolag2023,McAlpine2025}, indicating a shortcoming of the method. The implications on the calibration of the Local Universe Model are discussed in \cref{app:lum} while possible causes for this effect are proposed in \cref{sec:dis}.


\subsection{Massive galaxy clusters}

\paragraph{Detection rate} The detection rate for each cluster is shown in \cref{fig:drate}, for both the LUM and opt-LUM algorithms. Introduced in \cta{Pfeifer2023}, this quantifies the rate at which a match below a given \pval{} is found. The \pval{} thresholds cover the range from \sig{2} to \sig{4}, following the convention of \cta{McAlpine2025}. 

We first discuss the results of the LUM algorithm. More than half of the studied clusters (7/12) are found in more than 60\% of the simulations with (at least) \sig{2} deviation from randomness. Four of them reach a \sig{3} deviation in at least 20\% of the simulations. Marginally, the \sig{4} threshold is achieved for Shapley (A) and Norma. Fornax, a less massive but very nearby cluster, is recovered only in less than 20\% of the simulations. Ophiuchus, close to the Zone of Avoidance and thus poorly constrained, also performs below average with a 30\% detection rate at \sig{2}. Lastly, Hercules, Leo, Perseus and Coma display a detection rate between 60\% to 40\% at \sig{2} with no detection at \sig{3}, which is unexpected given their respective observed mass and the good data coverage in their respective surroundings. 

The opt-LUM shows significantly improved detection rates with respect to the standard LUM. All clusters have a detection rate higher than 80\% at \sig{2}, except Virgo, which scores slightly below this threshold and Ophiuchus, reaching only 60\%. The remarkable increase in detection rate for Coma, Perseus, Hercules and Leo is the subject of \cref{sec:dis:cope}. Half of the selected clusters achieve a \sig{3} deviation in more than 20\% of the simulations and most clusters (10/12) have a non-null \sig{4} detection rate. 

\paragraph{Separation} The separation between the clusters and their simulated counterparts is presented in \cref{fig:sep}. This separation is computed in redshift space, to keep consistency with the matching procedure. Three quantities are presented: the separation in 3D $(s)$, its projection on the line of sight $(s_r)$ and the remaining transverse distance $(s_t)$. This parametrization is motivated by the symmetry of the problem, as the restriction to the line of sight of the velocity data as well as the redshift space distortion necessitate a peculiar treatment of the radial direction. For each quantity, the separation between three pairs of objects are computed: LUM matched halos to observed clusters (blue), opt-LUM matched halos to observed clusters (red) and opt-LUM matched halos to their optimized target (green). 

We first discuss the results of the LUM algorithm. The three-dimensional separation roughly grows with distance, from less than $5\Mpch$ for Virgo and Fornax up to more than $20\Mpch$ for Hercules or Shapley (B). For the majority of the clusters, the scatter is more pronounced in the plane of the sky than in the radial direction (\ie{} $s_t / s_r > \sqrt{2}$), an effect that we attribute to the relatively low sensitivity of \emph{radial} velocities to the \emph{transversal} positions of objects due to projection effects. Our separations are consistent with \cta{Pfeifer2023}, yielding a slightly tighter scatter for clusters within $50\Mpch$. For Perseus, Coma, Leo and Ophiuchus, the recovered scatter is greater than that of \cta{Pfeifer2023}. Some of the reconstructed clusters align better with the 2MASS galaxy groups of \citet{Tully2015} than with the MCXC-II positions, this is the case for  Virgo, Norma, Coma, Ophiuchus, Shapley (A, B). This is expected, as the grouping of the CF4 velocity data is based on that of \citet{Tully2015}.

The opt-LUM approach always leads to separations of the observed clusters that are (at least) higher than those of the LUM. This is an expected result, as this method allows for the anchor to which the halos are matched to move. Three different behaviors emerge from the application of the opt-LUM algorithm. First, for Virgo, Hydra, Norma, Shapley (A \& B), the results are very similar to that of the LUM, indicating that there is no ambiguity on the group of halos that can be identified as counterparts. Second is the case of Fornax and Centaurus, for which the opt-LUM slightly shifts the target both in the plane of the sky. These differences of a few $\!\Mpc/h$ can be attributed to errors in the positions of the clusters and 
static biases in the resimulations, be they positional (on $d$) or dynamical (on $v_r$) as the redshift is a combination of both ($cz = v_r + H_0 d$).

Last, it appears that for Perseus, Leo, Coma, Ophiuchus and Hercules, the target is moved between $10-20\Mpch$ away from the cluster, almost uniquely in the transversal direction. In these five cases, the target converges to a cloud of massive halos scattered by less than $2-3\Mpch$ for Coma and Leo, $5-10\Mpch$ for Perseus, Ophiuchus and more than $10\Mpch$ for Hercules. This configuration is confirmed by visualization of the halos in 3D\footnote{This vizualization is available online \href{https://sketchfab.com/3d-models/cf4-nl-isos-mcxc-ii-clusters-mass-1e13-81ad65433bed4ca097f03ba9fc05a030}{under this link}.}. Furthermore, the simulated counterparts of Coma, Perseus and Hercules selected in SLOW as well as the so-called ``best match'' of \cta{Pfeifer2023} (not reported in \cref{fig:sep}]) appear to be part of these offset clouds. This result is further discussed in \cref{sec:dis:cmp}.

\paragraph{Mass} The posterior mass distribution of the identified counterparts is shown in \cref{fig:mass} for both LUM and opt-LUM methods. Mass estimates from other constrained simulations as well as from observations are overplotted. We note that a large scatter is found in the literature, across methods and sources. For instance, estimations from \citet{Tully2015} are systematically higher than others, indicating that the estimated mass may not be comparable with the $M_{500c}$ convention adopted in this paper.

The posterior mass distribution of halos selected by the LUM metric is consistent with the literature, with the notable exception of Perseus as well as Coma and Virgo to a lesser extent, who are less massive in this work than in others. This issue is solved when considering the masses of the opt-LUM counterparts, then, the mass estimations of all clusters are consistent with the literature. 

\subsection{Velocity field}

The application of the VFO metric on our data is presented in \cref{fig:vfo_wrt_hamlet}, using the current work as base for comparison. An extended version of this figure is found in the appendix \cref{app:sup_fig}. For consistency with the results published in \citet{Stiskalek2025}, the VFO is applied to only 20 out of the 100 simulations. We select the first and last state of each chain, so as to ensure that the tested simulations are not correlated (see \cref{app:mc_conv}). 

While the numerical values differ between test-catalogs, the ranking of the methods remain the same. \HamletPM{} is outperformed by both redshift-based constrained simulations, namely CSiBORG1 and Manticore-Local. Our method however scores better than the CLONES simulations, based on the Cosmicflows-2 data \citep{Tully2013} with the WF/RZA/CRs pipeline of the CLUES. 

Interestingly, the particle-mesh model does not grade as well as the linear one. This is confirmed by the fitted $\beta$ parameter, which indicates that the particle-mesh model over-estimates the amplitude of the velocity field by a factor two ($\beta^{-1} = 2.18 \pm 0.21$) while the linear model does not show this effect ($\beta^{-1} = 1.03 \pm 0.03$). As further discussed in \cref{sec:dis}, this is in all likelihood due to the systematic differences between the particle-mesh code used for the inference and the higher-resolution gravity solver used for the final simulations \citep{Stopyra2024}.

\section{Discussion}
\label{sec:dis}

This work extends the forward-modeling \Hamlet{} framework presented in \citet{Valade2022, Valade2023, Valade2024} to produce a novel method, dubbed \HamletPM{}, aimed at constraining initial conditions for cosmological simulations directly from velocity data. All in all, \HamletPM{} appears as a leap forward for the analysis of peculiar velocity data at field-level, compared to the traditional heterogeneous approach. While the traditional CLUES pipeline designed to produce such initial conditions is heterogeneous and limited by (1) linear theory and (2) the tension between the modeling of the observations by the Wiener Filter and their actual form \citep{Doumler2013, Pfeifer2023, Dolag2023}, our code \HamletPM{} is able to properly model non-linearities in the velocity fields with coarse particle-mesh dynamics \citep{Li2022} as well as complex observational effects. Moreover, \HamletPM{} is entirely Bayesian and thus allows for a self-consistent approach to the problem of generating constrained initial conditions. The full-forward modeling framework is flexible and can be easily extended or corrected. This opens the doors to further improvements of our method. 

In parallel, this work contributes to the effort of quantifying and standardizing tests for constrained simulations with the introduction of the opt-LUM, an iterative simulated counterparts matching algorithm built on the methodology of the LUM that captures static biases in the posterior distribution of counterparts. Furthermore, the opt-LUM appears to yield better mass estimations than the simple LUM.

\subsection{Tensions with the underlying model}

The treatment of peculiar velocity data is admittedly challenging, as measurements are scarce, noisy and prone to interpretation biases. While \HamletPM{} paves the way to the building of a realistic observational model for such data, shortcomings are still to be acknowledged. \HamletPM{} over-estimates the number of massive clusters above $10^{14}\Msh$ in the local universe, reaching +150\% for $10^{15}\Msh$ halos, which is not only at odds with \LCDM{} but also Local Universe estimates \citep{Bohringer2020, Dolag2023, McAlpine2025}. Compared to \LCDM{} predictions, more particles appear to be found in halos as there is no under-abundance of low-mass halos. Furthermore, the power spectrum of the initial conditions appears to be slightly over-estimated across the probed frequency range (by 5\% to 10\%), as does the amplitude of the evolved velocity field (by almost 220\%). 

Our results are very similar to those described by \citet{Stopyra2024} in which the authors notably discuss the over-abundance of massive halos in simulations constrained by the model presented in \citet{Jasche2019}, such as the CSiBORG1 constrained simulations suite \citep{Bartlett2021}. They conclude that this effect is due to the use of a 10-step particle-mesh algorithm in the inference loop and demonstrate that the issue can be solved with a use of a 20-step COLA gravity solver \citep{Tassev2013}, or alternatively by increasing the number of particle-mesh steps by an order of magnitude. 

\citet{Stopyra2024} suggests that, as the particle-mesh solver is not able to reproduce accurately densely collapsed objects, the method must exaggerate the contrast in the ICs in order to fit the observed galaxy density. Once resimulated with an accurate gravity solver, the over-estimation of the contrast in the ICs results in an over-abundance of massive halos. Our result here is slightly different, as we reconstruct the velocity field instead of the density field. In all likelihood, this demonstrates that the particle-mesh algorithm tends to under-estimate the amplitude of the late-time velocity fields in dense regions, forcing the ICs to deviate from the cosmological model and leading to biased resimulations. This would explain why the bias is strong for high-mass halos, where the velocity of the collapse is largely under-estimated by the particle-mesh algorithm and almost non-existant for low-mass halos, where the dynamics are properly resolved by the approximated gravity solver. 

We note that other effects may also contribute to the over-estimation of the contrast, such as the inhomogeneous Malmquist bias \citep{Boruah2022} or the mishandling of the selection function \citep[see fig. 9 of ][]{Carreres2023}. 

The over-abundance of massive halos may be problematic for the LUM-related metrics as an over-estimation of the mass of the halos matched with observational clusters would artificially inflate our performance. Yet, it is noted that the masses of the studied clusters are in good agreement with the literature, indicating that our LUM-results appear to not be boosted by an over-estimation of the mass.

\subsection{Comparison with other works}
\label{sec:dis:cmp}

The product of our work is twofold, as it offers (1) a picture of the density and velocity fields in the late-time Local Universe and (2) a series of cosmological simulations tailored to reproduce velocity observations. We thus compare our results not only to other constrained simulations but also to ``direct'' reconstructions of the Local Universe. We first start the discussion by a rapid qualitative overview, moving toward a more rigorous quantitative analysis. 

\paragraph{Qualitative comparison} Compared to other velocity-based reconstructions, the improvements appear in the increased sharpness of the reconstructed large-scale structure, with respect to previously published works \citep[\eg][]{Hoffman2018, Pfeifer2023, Sorce2023, Courtois2023, Valade2024}. We note the clear emergence of non-linear structures, such as thin walls and filaments. A similar conclusions arises from the comparison with direct reconstruction of the density field from redshift data, such as \citet{Carrick2015} or \citet{Lilow2024}. Yet, Manticore-Local, the most recent suite of redshift-based constrained simulations, offers a more precise picture of the cosmic web \citep[see fig. 9 and D3 of][]{McAlpine2025}.

\paragraph{Quantitative comparison} To complement the limitations of qualitative comparison, two metrics have been recently proposed. Both highlight the superiority of our approach over other velocity-based pipelines, while also demonstrating its shortcomings when compared against redshift-based works.

The VFO enables a comparison between constrained simulations and direct reconstructions from the angle of the obtained velocity field(s). The results, presented in \cref{fig:vfo_wrt_hamlet,fig:vfo_app}, show that the VFO ranks both our linear and non-linear results above other tested velocity-based reconstructions, considerably reducing the gap to redshift-based methods. For the CF4-$i$ data set, our linear reconstruction overtakes CSiBORG1 -- a former release of simulations constrained by the BORG algorithm, now replaced by Manticore-Local -- to match with \citet{Carrick2015, Lilow2024}. However we note that, in this case, our input data is related to the test data.

A similar conclusion is drawn from the application of the LUM metric to various (suites of) constrained simulations. Presented in \cref{fig:drate_cmp_rss} for two detection rates, the results evidence that \HamletPM{} recovers massive clusters of galaxies with a higher rate than \cta{Pfeifer2023} but a lower detection rate than Manticore-Local for most clusters. We note that Perseus and Coma are found at the same detection rate by \cta{Pfeifer2023}, which is further discussed in \cref{sec:dis:cope}. 

Comparison with the SLOW simulation is made difficult by the absence of statistical ensemble, as only one set of ICs was selected out of nine pre-run constrained simulations \citep{Sorce2018,Dolag2023}. Yet, we tested that 96\% (\resp{} 20\%) of our simulations have more than three counterparts with \pval{} $<0.01$ (\resp{} $0.0014$), ensuring that at least one in nine simulation would achieve a similar result. Six of our simulations find eight counterparts out for the twelve targeted clusters.

\paragraph{Conclusion}

The quantitative comparison of velocity-based reconstructions, supported by a qualitative assessment shows that our method outdoes previous works. This tends to indicate that the full forward-modeling is able to put tighter constraints on galaxy clusters than the standard CLUES pipeline. An sensitive improvement may also stem from the use of a larger data set, namely CF4 versus CF2 for SLOW and CF3 for \cta{Pfeifer2023}. 

Yet, our method is largely outperformed by Manticore-Local. We propose two explanations for this effect. First, the replacement of the  particle-mesh gravity solver to COLA has largely improved their results while reducing biases \citep{Stopyra2024, McAlpine2025}. Furthermore, Manticore-Local has a higher grid resolution ($3.9\Mpc$ vs. $5.22\Mpc$ here). More importantly, the reconstruction of Manticore-Local builds on redshift data, which is denser ($\sim$69\,000 galaxies vs. $\sim$31\,000 here) and orders of magnitude more precise (a median signal to noise of less than 1\% vs more than 180\% here). This observational comparative advantage of redshift data over velocity data should hold with upcoming surveys. 

Last but not least, reconstruction from redshift data has a clear advantage from the view point of cluster-based metrics such as the LUM. Indeed, the exact position of massive clusters is very well defined in the redshift data, while it is much less clear in the velocity data. From a pure geometrical point of view, radial velocities are weakly sensitive to transverse shifts in the positions of the galaxy clusters. The flattening of the counterparts' scatter around their associated observational clusters is symptomatic of this limitation. 

In parallel to the application to upcoming large new datasets, future works should focus on handling better observational systematics, such as the selection function or a better modeling of uncertainties stemming from the cross-calibration process; as well as improving the modeling of the density and velocity fields so as to probe deeper the non-linear regime. 

\section*{Acknowledgements}

AV thanks S. McAlpine as well as E. Hernandez-Martinez for sharing results of their respective works. AV has been supported by the Agence Nationale de la Recherche of the French government
through the program ANR-21-CE31-0016-03. RS acknowledges financial support from STFC Grant No. ST/X508664/1, the Snell Exhibition of Balliol College, Oxford.

\section*{Authors contributions}

AV designed and ran the constraining of the initial conditions, analyzed the data and wrote the manuscript. NL ran the cosmological simulations. DP did the 3D cosmography (Fig 3 and D1). RS ran and analyzed the VFO metric (Fig 4 and C1). NL, DP, RS, YH, SG and BT were part of the discussions and corrected the manuscript. 

\bibliographystyle{bibtex/aa}
\bibliography{main}

\appendix

\section{More on the forward model} 

\label{app:model}

\begin{figure*}
    \begin{center}
        \begin{tikzpicture}
            [
            par/.style={
                ellipse,
                double,
                draw=black,
                color=black,
            },
            rec/.style={
                ellipse,
                draw=black,
                color=black,
            },
            data/.style={
                rectangle,
                draw=black,
                color=black,
            },
            xscale=2, 
            yscale=.75,
            ]
            \node [par]  (dist)  at (1, 6) {$d$};
            \node [rec]  (zcos)  at (3, 5) {$z_{\rm cos}$};
            \node [rec]  (vr)    at (3, 4) {$v_r(\bm{r})$};
            \node [data] (hr)    at (2, 4) {$\hat{\bm{r}}$};
            \node [rec]  (z)     at (4, 4) {$z$};
            \node [data] (zobs)  at (5, 4) {$z^{\rm o}$};
            \node [rec]  (dl)    at (4, 6) {$d_L$};
            \node [data] (vcmb)  at (4, 8) {$\bm{v}_{\rm CMB}$};
            \node [data] (hr2)  at (4, 5) {$\hat{\bm{r}}$};
            \node [rec]  (mu)    at (5, 6) {$\mu$};
            \node [data] (muobs)  at (6, 6) {$\mu^{\rm o}$};

            \node [par] (delini)   at (1, 1) {$\tilde{\delta}^{\rm ini}$};
            \node [rec] (vcic)  at (2, 2) {$\bm{v}^{\rm CiC}$};
            \node [rec] (rhocic)  at (2, 0) {$\rho^{\rm CiC}$};
            \node [rec] (vlin)  at (3, 0) {$\bm{v}^{\rm lin}$};
            \node [rec] (vhyb)  at (3, 2) {$\bm{v}^{\rm hyb}$};

            \draw[draw=gray, dotted, thick] (0.5, -0.75) rectangle  +(3, 3.5);
            \node[font=\tiny] (plate_m) at (1.05, -.25) {$\times ~ m$ grid cells};
            \draw[draw=gray, dotted, thick] (0.5, 3) rectangle  +(6, 3.75);
            \node[font=\tiny] (plate_n) at (6, 3.25) {$\times ~ n$ observations};

            \path (dist)   edge [->] (zcos)
                  (zcos)   edge [->] (dl)
                  (dl)     edge [->] (mu)
                  (dist)   edge [->] (dl)
                  (dist)   edge [->] (vr)
                  (hr)     edge [->] (vr)
                  (zcos)   edge [->] (z)
                  (vr)     edge [->] (z)
                  (delini)  edge [->] (rhocic)
                  (delini)  edge [->] (vcic)
                  (vcic)  edge [->] (vhyb)
                  (vlin)  edge [->] (vhyb)
                  (rhocic)  edge [->] (vhyb)
                  (rhocic)  edge [->] (vlin)
                  (vhyb) edge [->] (vr)
                  (vcmb)   edge [->] (dl)
                  (hr2)   edge [->] (dl);

            \path (mu) edge [<->, double] (muobs)
                  (z)  edge [<->, double] (zobs);

            \node (lgd) at (9, 7.) {legend};
            \draw[draw=black] (7.75, -.25) rectangle +(1.5, 7);

            \node [font=\tiny] (lgd_cpt) at (8.5, 6.35) {compute};
            \path (8.25, 6) edge [->] (8.75, 6);
            \node [font=\tiny] (lgd_lkl) at (8.5, 5.35) {likelihood};
            \path (8.25, 5) edge [<->, double] (8.75, 5);
            \node [par, font=\tiny]  (leg_par) at (8.5, 4) {parameter};
            \node [rec, font=\tiny]  (leg_rec) at (8.5, 3) {internal};
            \node [data, font=\tiny] (leg_dat) at (8.5, 2) {data};
            \draw[draw=gray, dotted, thick] (8, 0) rectangle  +(1, 1);
            \node[font=\tiny] (leg_plate) at (8.5, 0.5) {plate};
        \end{tikzpicture}
    \end{center}
    \caption{Directed Acyclic Graph (DAG) of the forward model. Each physical quantity is represented as node of the graph -- either in a ellipse for modeled quantities or in a square for observational data. Free parameters of the posterior are on the left-most side of the diagram and highlighted by a double ellipse. Simple arrows indicate a relationship between quantities (\ie{} ``A is computed from B''), while double arrows indicate a comparison (\ie{} a likelihood function). Note that prior probability functions do not appear in this representation. Dashed square boxes specify the dimensionality of the physical quantities. The definitions and
    relationships between quantities are found in \cref{sec:obs_model}.}
\label{fig:dag}
\end{figure*}
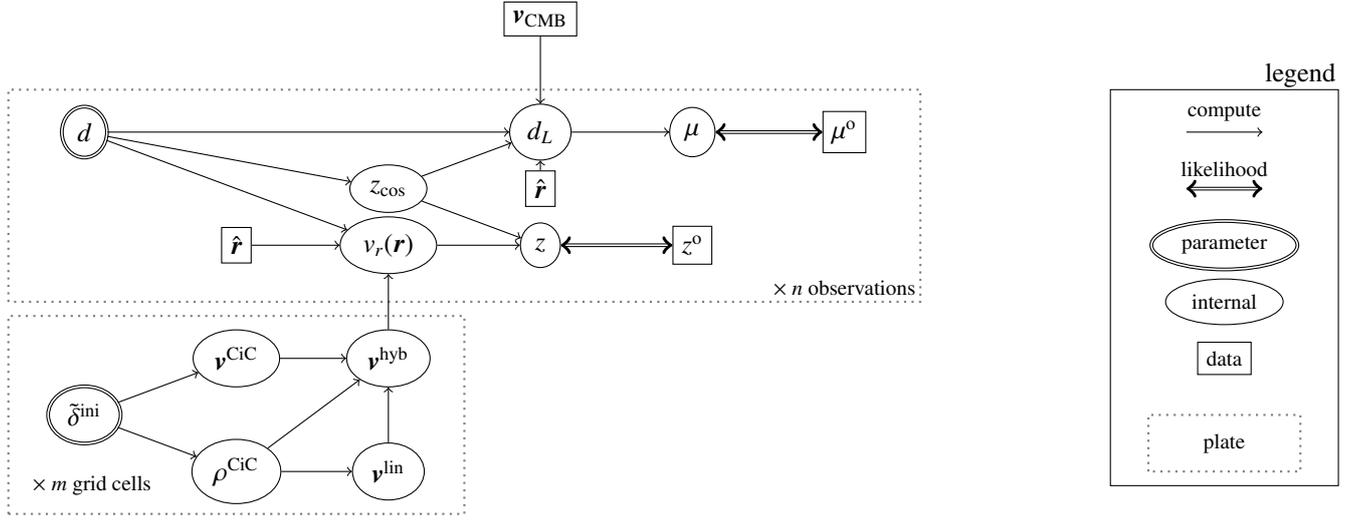

In a Bayesian framework, selections effects are written under the form $P(\lO \mid {\rm obs})$, where $\lO$ is the event that ``is-observed'' and ${\rm obs}$ is any observable quantity. Because Cosmicflows-4 is a compilation of several surveys, its selection function $P(\lO \mid {\rm obs})$ is composite, and its description requires the modeling of other survey- and method-dependent observables, such as the apparent magnitude in different bands. 

To circumvent the issue, we note that these observables ${\rm obs}$ are constructed from the free parameters, and that the selection functions can be seen as functions of the free parameters. In our case, the selection functions can thus be written as $P(\lO \mid d, \Delta)$, which we simplify to $P(\lO \mid d)$, assuming the effects of the velocity field on the selection is negligible. 

We then note that the posterior depends on the product of the selection function $P(\lO \mid d)$ and the prior $P(d)$ such that
\begin{equation}
    P(d, \Delta \mid \lM, \lZ) \propto P(\lO \mid d) \times P(d) \propto P(d \mid \lO).
\end{equation}
The right-hand side quantity, $P(d \mid \lO)$, is in fact the distribution of \emph{true} distances in our observational catalog, which can be approximated by 
\begin{equation}
    P(d \mid \lO) = \big(\lN(0, \sigma_{v, \Lambda{\rm CDM}}) \ast \lH_{d_z}\big)(d)
\end{equation}
where $\sigma_{v, \Lambda{\rm CDM}}\approx 300 \kms$ is the expected distribution of peculiar velocities in \LCDM, the asterisk $\ast$ denotes a convolution, and $\lH_{d_z}$ is the distribution of redshift-derived distances
\begin{equation}
    d_z = \frac{H_0}{c} \int_0^z \frac{{\rm d} y}{\Omega_m (1 + y)^3 + \Omega_\Lambda}.
\end{equation}

We apply separately this method to  extract (1) groups with a majority of SDSS-PV measurements, (2) groups with a majority of 6dF-PV measurements and (3) other groups. For each sub-catalog, we describe the selection function as
\begin{equation}
    P(\lO \mid d) P(d) = P(d \mid \lO) = \lH_{d_z}(d)
\end{equation}
where $\lH_{d_z}$ is the distribution of observed redshift distances computed from the observed redshifts.




\section{Convergence of the Monte Carlo chain}
\label{app:mc_conv}

Given a Bayesian model, a Monte Carlo technique aims to sample the posterior distribution resulting in a statistical sample of $n$ (vectors of) parameters $\big\{\theta_i\big\}$, such that $\theta_i \sim P(\theta_i \mid D)$ where $D$ is the data. Ideally, consecutive states of the Monte Carlo chains are independent, \ie{} $P(\theta_i \mid \theta_{i - 1}, D) = P(\theta_i \mid D)$. Yet, as the Monte Carlo sampling explores the parameter space by constructing each state  from the previous one, correlation along the chain may occur. In the presence of internal correlation, the number of statistically independent samples is lower than the length of the Monte Carlo chain. To avoid giving too much weight to volumes of the parameters space that are over-sampled, the chain must be ''thinned out'' until consecutive states are uncorrelated. By construction, there is no correlation between separated Monte Carlo chains. 

\begin{figure}
    \begin{center}
        \includegraphics[width=.5\textwidth]{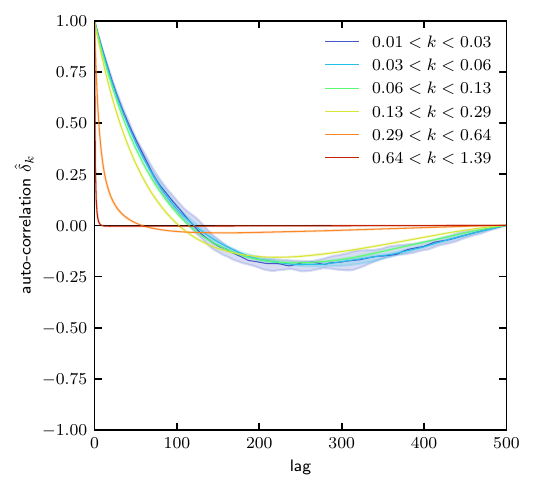}
    \end{center}
    \caption{Autocorrelation of the (real part) of the Fourier modes of the initial density field, binned in frequencies. The shades represent the scatter alongside the ten run Monte Carlo chains.}
    \label{fig:chain_autocorr}
\end{figure}

\Cref{fig:chain_autocorr} presents the autocorrelation of the chain of the modes of the over-density field, binned in frequencies. It appears that the correlation between high-frequencies modes very rapidly decreases while low-frequencies modes ($k < 0.13\hMpc$, \ie{} $\lambda > 50\Mpch$) are correlated over about 100 Monte Carlo steps. 

The initial conditions selected for the 100 simulations presented in this work are extracted from each of the 10 Monte Carlo chains with a thinning of 50 steps. This means that consecutive simulations may not be entirely independent on the large scales, yet, note that (1) the power added to increase the resolution of the initial conditions is entirely random and that (2) this paper essentially focuses on dark matter halos, whose properties are affected by high-frequency modes \citep{Sawala2021}. 

\section{Local Universe Model}
\label{app:lum}

In each simulation, each halo can be matched to only one observational cluster. The optimal configuration is computed by optimizing the sum of the \pval{}s through the ``Hungarian algorithm''  \citep{Kuhn1955} . Matches with $p_{\rm val}>0.023$ (\sig{2}) are discarded, meaning that all clusters may not have a counterpart in all simulations. 
 
Contrary to \cta{Pfeifer2023} and \cta{McAlpine2025}, we argue in favor of running the LUM in redshift space (\ie{} setting the halos at their redshift position). This choice is motivated by the use of redshift positions for the observational clusters, which constitutes the bulk of the values found in the literature.

The \pval{}s computed by the LUM are cosmology-dependent. This appears in the pseudo-density, mass-dependent parameter $\alpha(M)$ defined as 
\begin{align}
  &p_{\rm val}(r, M) = \int_0^s P(s, M) {\rm d} s \\
  &P(r, M) = e^{- \frac{4}{3} \pi \alpha(M) r^3 } - e^{- \frac{4}{3} \pi \alpha(M) (r + \Delta r)^3 }
\end{align}
where $P(r, M)$ encodes the probability to find a halo of mass $>M$ within a distance $[r, r + \Delta r]$. The form of $P(r, M)$ is exact for unclustered halos where $\alpha(M)$ is then simply the density of halos of mass $>M$. To account for clustering, the $\alpha$ parameter can be calibrated on cosmological simulations. The best practice is then to use an accompanying series of random simulations sharing the cosmology of the studied constrained simulation. This is the path chosen by \cta{Pfeifer2023} and \cta{McAlpine2025}. Alternatively, one may calibrate the LUM by choosing random observers in the studied constrained simulation \cpa{Hernandez-Martinez2024} . As it creates the risk of biasing the random signal, this approach should be followed only after testing that the constrained simulation is statistically consistent with the underlying cosmology. 

For this work, we have chosen to use the available computational power to build a large enough statistical sample of constrained simulations. In absence of companion random simulations, the first approach to derive the LUM cannot be used. Furthermore, as presented below in \cref{fig:hmf} and discussed in \cref{sec:res:lcdm,sec:dis}, our simulations deviate from the expected halo mass distribution, making the second approach inapplicable. To account for this effect, we use the calibration of \cta{Pfeifer2023} that we correct as such
\begin{equation}
  \alpha(M) = \alpha_{\rm P23}(M) \frac{\int_{m > M} \rho_{\rm cst}(m) {\rm d} m}{\int_{m > M} \rho_{\rm P23}(m) {\rm d} m}
\end{equation} 
where $\rho_{\rm cst}(M)$ is the halo mass function in the constrained volume ($d < 170\Mpch$) and $\rho_{\rm P23}(M)$ is the halo mass function predicted by \citet{Tinker2008} for the cosmology of \cta{Pfeifer2023}. Physically, this correction only accounts for the variation of halo mass function with cosmology while neglecting the difference in clustering. In effect, it increases the pseudo-density for high-mass halos, from $+ 5\%$ at $M_{200c} = 10^{14}\Msh$ up to $+150\%$ at $M_{200c} = 10^{15} \Msh$. For a given halo-cluster configuration, our cosmology predicts a higher \pval{} than that of \cta{Pfeifer2023}.  

\subsection{Shift of Coma and Perseus in velocity reconstructions}
\label{sec:dis:cope}

\begin{figure*}
    \begin{center}
        \includegraphics[height=.3\textheight]{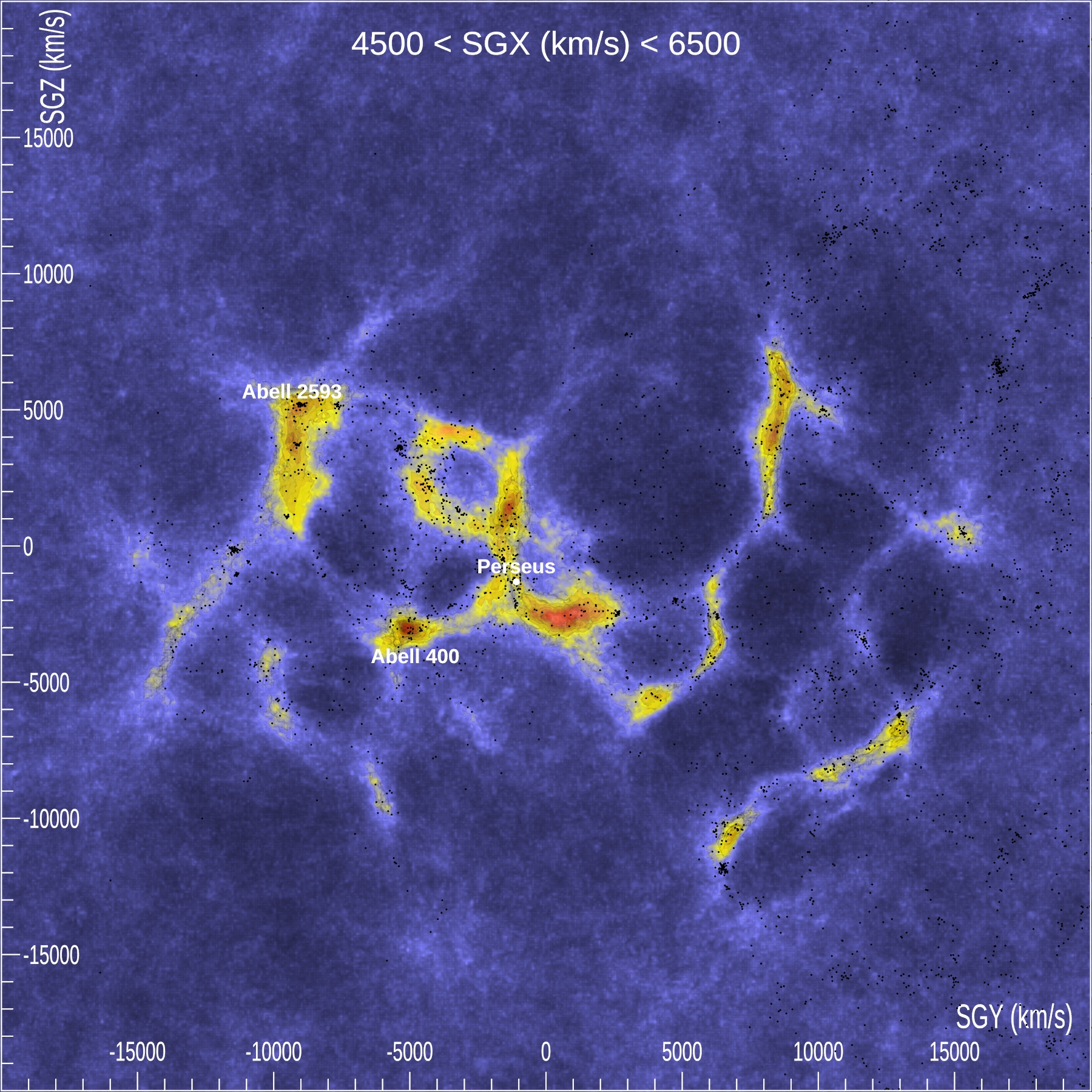}  
        \includegraphics[height=.3\textheight]{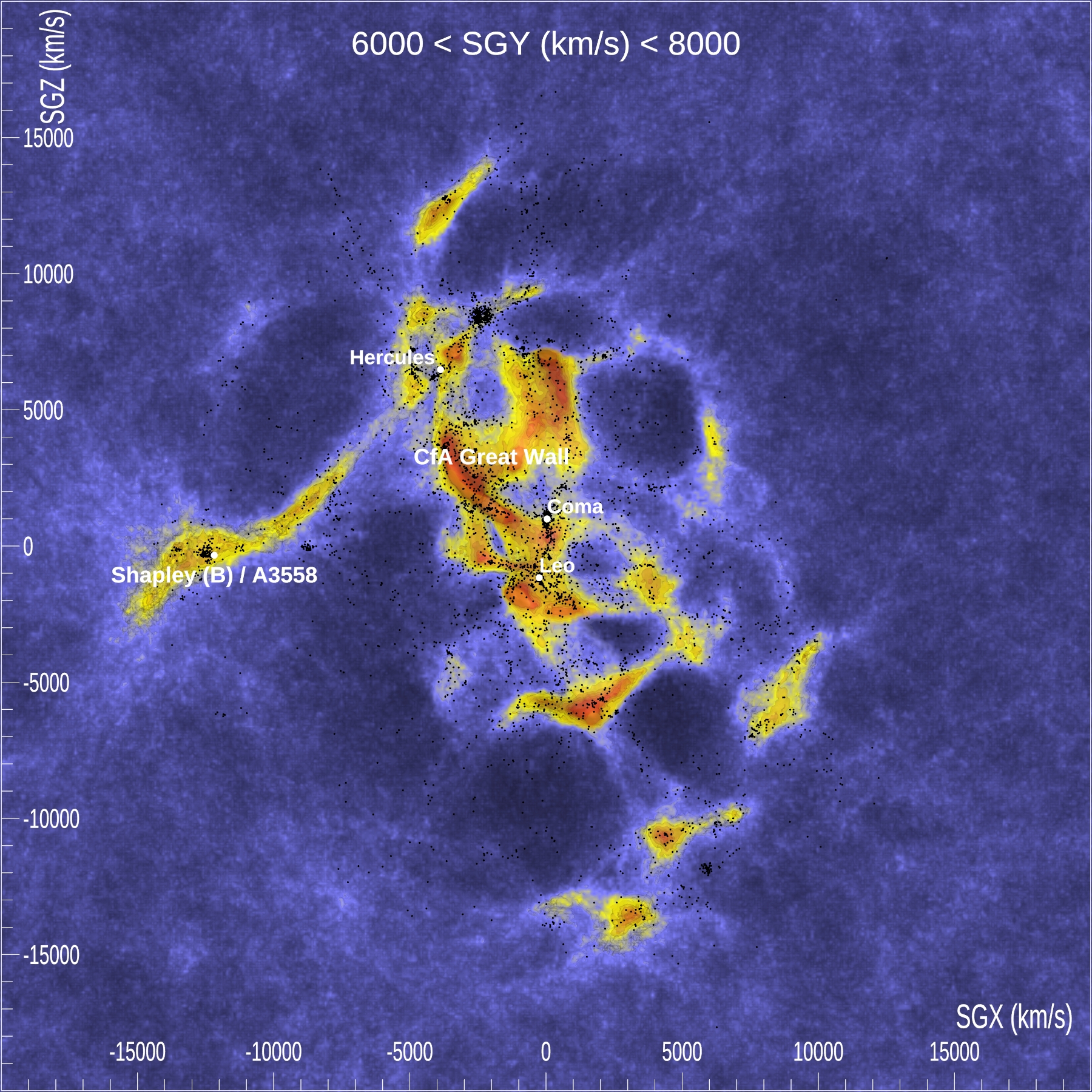}  
    \end{center}
    \caption{Density field around the regions of Coma and Perseus. These slices have been selected to show the observational positions of the clusters as well as the density peaks to which the opt-LUM counterparts belong.}
    \label{fig:cope-slices}
\end{figure*}

The position on the sky of the simulated Coma and Perseus is very interesting. The opt-LUM algorithm proposed here enables us to properly quantify for the first time this shift, which seems to have been present in the literature for a decade. Indeed, all the previously studied velocity-based constrained simulations appear to systematically shift these two clusters by a solid $10-20\Mpch$ orthogonally to the line of sight. Furthermore, all linear reconstructions \citep{Sorce2016, Graziani2019, Sorce2023, Hoffman2024, Valade2024} as well as the quasi linear fields from \citet{Hoffman2018} and \cta{Pfeifer2023}, present a density peak about $10\Mpch$ from Perseus' position in +SGY direction, within the zone of avoidance. This is also the case for Coma in our reconstruction. However, the shift in peak density out of the super-galactic planes canonically used to visualize the density field, which prevents us from verifying easily if this is also the case in the literature. We highlight that such a shift appears not to be present in redshift-based reconstructions such as Manticore-Local \cpa{McAlpine2025}, which is consistent with the fact that the redshift distribution is in this case the only source of information. 

Furthermore, none of the three studied simulation suites (\cta{Pfeifer2023}, \cta{Hernandez-Martinez2024} and \HamletPM) uses the same pipeline nor the same data source. 
While we note that these data sets build up on each other and thus share constraints, the addition of new, independent additional data should have mitigated systematic biases. The contribution of new data is significant in these regions, with the addition of the SDSS-PV sample for Coma and the CF4-TF sample for Perseus. Similarly, the inter-calibration between sources catalogs is rerun between successive Cosmicflows releases, making it unlikely to be the explanation for this shift. 

The shifted position of Perseus is in the ZoA and appears to be close to 3C129.1, three times lighter than Perseus itself \citep{Sadibekova2024}. The shifted Coma is found in the CfA Great Wall, where no other massive cluster is found in the MCXC-II catalog. The density field in these regions is presented in \cref{app:sup_fig}.

If confirmed, this result may lead to two separate conclusions. Either velocity data is able to detect a major issue in our physical model, such as a strong decoupling between the presence of galaxies and the dark-matter density field or in the laws of gravitation. Alternatively, it may show that velocity-based reconstructions have to be checked carefully for spurious detections. We leave this open problem to the community.

\section{Supplementary material}
\label{app:sup_fig}

\begin{figure*}
    \begin{center}
        \includegraphics[height=.4\textheight]{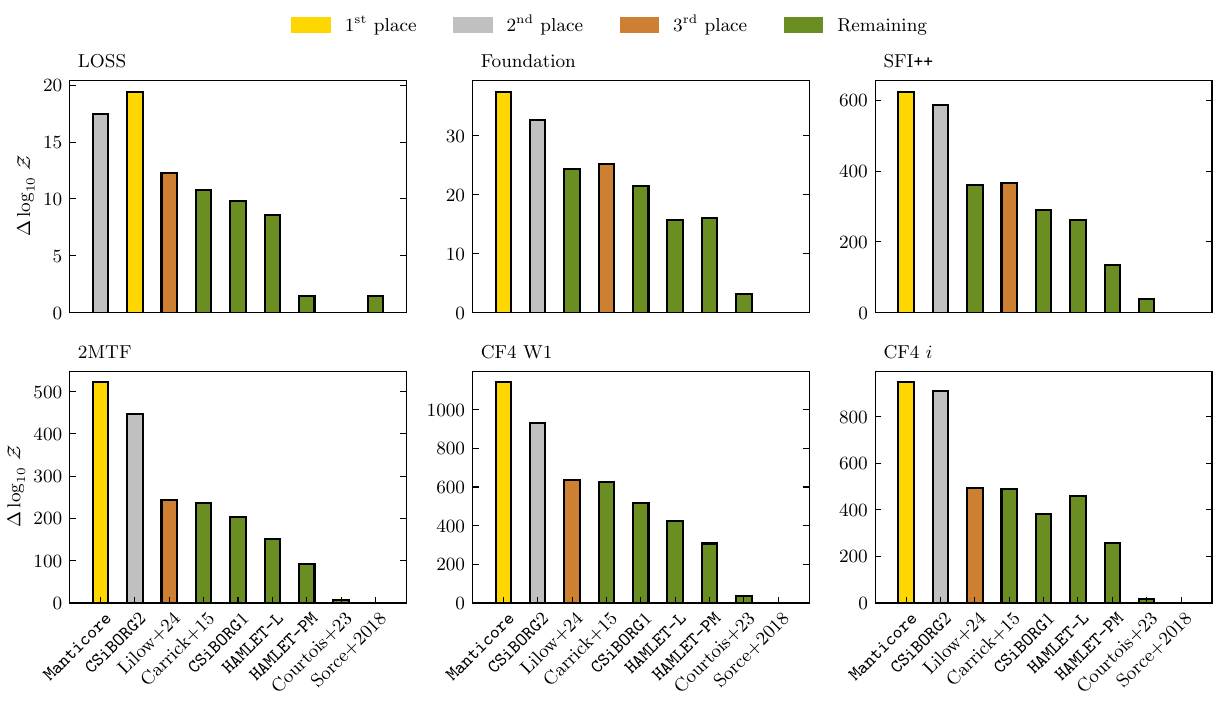}
    \end{center}
    \caption{The results of the original Velocity Fields Olympics are presented here, augmented by our linear and non linear reconstructions \citep[][this work]{Valade2024}, as well as the last constrained simulations of the AQUILA consortium, namely Manticore-Local \citep{McAlpine2025}. Our methods tighten the gap between velocity-based and redshift-based reconstructions by performing markedly better than all other tested velocity-based  methods while remaining below the scores of redshift-based reconstructions. }
    \label{fig:vfo_app}
\end{figure*}

Supplementary material is gathered in this section. It includes an extended visualization of the VFO results in \cref{fig:vfo_app}, as well as the tables gathering the posterior information on massive galaxy clusters in \cref{tab:LUM_matched} (LUM-matched) and \cref{tab:opt-LUM_matched} (opt-LUM-matched).

\begin{table*}
  \caption{Summary of the LUM matched halos}\label{tab:LUM_matched}
  \begin{center}
    \begin{tabular}{l|ccc|cc|c|c}
    Cluster       & SGX                            & SGY    & SGZ    & $M_{200c}$ & $M_{500c}$ & \pval{} & separation           \\ 
    ~             & \multicolumn{3}{c|}{[Mpc/$h$]}                   & \multicolumn{2}{c|}{[$10^{14}\,M_{\odot}/h$]} & $10^{-3}$ & [Mpc/$h$] \\ 
    \hline
    Virgo & $-2.3$ & $12.3$ & $0.2$ & $1.2^{+0.8}_{-0.6}$ & $0.8^{+0.6}_{-0.5}$ & $6.6^{+5.1}_{-2.9}$ & $3.2^{+0.7}_{-0.7}$ \\
    Fornax & $0.0$ & $-10.1$ & $-8.8$ & $1.3^{+0.5}_{-0.4}$ & $0.9^{+0.5}_{-0.3}$ & $8.6^{+8.7}_{-6.0}$ & $3.2^{+2.1}_{-1.2}$ \\
    Centaurus & $-28.2$ & $13.0$ & $-11.1$ & $4.1^{+2.2}_{-2.7}$ & $2.6^{+1.5}_{-1.7}$ & $4.0^{+6.1}_{-2.6}$ & $5.7^{+2.9}_{-2.3}$ \\
    Hydra & $-23.3$ & $22.4$ & $-22.0$ & $3.1^{+1.5}_{-1.2}$ & $2.1^{+1.2}_{-0.9}$ & $2.4^{+5.6}_{-1.9}$ & $3.9^{+1.6}_{-1.4}$ \\
    Norma & $-46.5$ & $-5.6$ & $6.6$ & $4.0^{+3.4}_{-2.2}$ & $2.7^{+2.4}_{-1.5}$ & $1.0^{+5.3}_{-0.9}$ & $3.8^{+2.6}_{-1.5}$ \\
    Perseus & $52.0$ & $-10.0$ & $-15.0$ & $2.5^{+3.8}_{-1.4}$ & $1.7^{+2.6}_{-0.9}$ & $8.8^{+9.0}_{-6.6}$ & $4.8^{+8.0}_{-2.1}$ \\
    Leo & $-7.3$ & $63.4$ & $-15.2$ & $4.7^{+2.0}_{-2.2}$ & $3.2^{+1.2}_{-1.5}$ & $10.9^{+5.7}_{-7.1}$ & $10.9^{+1.3}_{-5.9}$ \\
    Coma & $-2.8$ & $69.7$ & $8.6$ & $2.9^{+4.8}_{-1.7}$ & $1.9^{+2.9}_{-1.1}$ & $9.8^{+9.0}_{-5.5}$ & $6.2^{+12.8}_{-3.0}$ \\
    Ophiuchus & $-54.3$ & $11.5$ & $62.0$ & $4.9^{+4.1}_{-2.2}$ & $3.4^{+2.7}_{-1.5}$ & $10.4^{+8.7}_{-8.5}$ & $9.4^{+12.2}_{-3.0}$ \\
    Hercules & $10.9$ & $54.9$ & $74.5$ & $5.1^{+4.4}_{-3.1}$ & $3.5^{+2.8}_{-2.1}$ & $6.0^{+6.7}_{-3.9}$ & $7.9^{+11.2}_{-4.1}$ \\
    Shapley (A) & $-104.5$ & $55.7$ & $6.9$ & $6.4^{+3.3}_{-3.6}$ & $4.4^{+2.0}_{-2.6}$ & $2.7^{+8.2}_{-2.5}$ & $7.5^{+5.4}_{-4.0}$ \\
    Shapley (B) & $-124.8$ & $73.7$ & $-3.4$ & $7.6^{+2.6}_{-3.9}$ & $4.8^{+1.9}_{-2.3}$ & $3.3^{+9.4}_{-2.6}$ & $13.6^{+16.4}_{-9.1}$ \\
    \end{tabular}
  \end{center}
\end{table*}

\begin{table*}
  \caption{Summary of the opt-LUM matched halos}\label{tab:opt-LUM_matched}
  \begin{center}
    \begin{tabular}{l|ccc|cc|c|c}
    Cluster       & SGX                            & SGY    & SGZ    & $M_{200c}$ & $M_{500c}$ & \pval{} & separation           \\ 
    ~             & \multicolumn{3}{c|}{[Mpc/$h$]}                   & \multicolumn{2}{c|}{[$10^{14}\,M_{\odot}/h$]} & $10^{-3}$ & [Mpc/$h$] \\ 
    \hline
    Virgo & $-5.8$ & $13.5$ & $-0.1$ & $1.7^{+0.7}_{-0.5}$ & $1.2^{+0.5}_{-0.4}$ & $1.5^{+5.6}_{-1.3}$ & $5.3^{+0.9}_{-1.1}$ \\
    Fornax & $3.5$ & $-12.4$ & $-6.3$ & $1.2^{+0.5}_{-0.5}$ & $0.8^{+0.3}_{-0.3}$ & $2.0^{+6.0}_{-1.5}$ & $6.7^{+1.1}_{-1.1}$ \\
    Centaurus & $-25.3$ & $12.0$ & $-15.1$ & $3.8^{+2.4}_{-1.8}$ & $2.7^{+1.5}_{-1.3}$ & $1.1^{+4.1}_{-1.0}$ & $9.8^{+5.3}_{-3.4}$ \\
    Hydra & $-23.5$ & $23.2$ & $-21.8$ & $2.9^{+1.6}_{-1.1}$ & $2.1^{+1.1}_{-0.8}$ & $0.3^{+1.8}_{-0.3}$ & $4.0^{+1.5}_{-1.2}$ \\
    Norma & $-46.4$ & $-5.1$ & $7.1$ & $4.2^{+3.6}_{-2.3}$ & $2.8^{+2.5}_{-1.5}$ & $1.1^{+6.7}_{-1.0}$ & $4.0^{+3.2}_{-1.5}$ \\
    Perseus & $54.6$ & $5.9$ & $-25.4$ & $5.3^{+2.6}_{-2.8}$ & $3.5^{+1.9}_{-1.8}$ & $0.6^{+6.5}_{-0.5}$ & $21.6^{+3.1}_{-3.2}$ \\
    Leo & $-11.6$ & $60.9$ & $-18.4$ & $3.6^{+2.0}_{-1.9}$ & $2.5^{+1.3}_{-1.3}$ & $0.2^{+2.6}_{-0.2}$ & $11.7^{+1.2}_{-1.3}$ \\
    Coma & $-22.6$ & $65.6$ & $15.7$ & $4.8^{+2.1}_{-1.5}$ & $3.3^{+1.4}_{-1.1}$ & $0.2^{+0.7}_{-0.2}$ & $23.7^{+2.5}_{-2.2}$ \\
    Ophiuchus & $-49.0$ & $18.6$ & $65.3$ & $4.1^{+4.4}_{-2.6}$ & $2.9^{+2.9}_{-1.9}$ & $1.7^{+10.7}_{-1.5}$ & $18.3^{+3.6}_{-4.2}$ \\
    Hercules & $-2.1$ & $60.4$ & $74.4$ & $4.3^{+5.0}_{-3.1}$ & $3.0^{+3.0}_{-2.1}$ & $3.1^{+8.4}_{-2.8}$ & $19.5^{+3.9}_{-2.6}$ \\
    Shapley (A) & $-106.3$ & $57.5$ & $7.1$ & $7.7^{+2.5}_{-3.5}$ & $5.2^{+1.8}_{-2.7}$ & $2.8^{+7.7}_{-2.4}$ & $8.6^{+14.3}_{-3.2}$ \\
    Shapley (B) & $-129.7$ & $72.2$ & $-4.3$ & $5.8^{+3.4}_{-4.0}$ & $3.8^{+2.3}_{-2.8}$ & $2.9^{+5.9}_{-2.5}$ & $10.8^{+7.6}_{-4.2}$ \\
    \end{tabular}
  \end{center}
\end{table*}

\end{document}